\documentclass[onecolumn,prd,aps]{revtex4}

\usepackage{amsmath}
\usepackage{verbatim}
\usepackage{graphicx}
\usepackage[usenames]{color}
\usepackage{psfrag}

\usepackage{hyperref}
\usepackage{amssymb}
\usepackage{enumerate} 

\newcommand{\beq}{\begin{equation}}
\newcommand{\eeq}{\end{equation}}
\newcommand{\beqa}{\begin{eqnarray}}
\newcommand{\eeqa}{\end{eqnarray}}

\newcommand{\mpl}{M_{\rm pl}} 
\newcommand{\geff}{G_{\rm eff}}
\newcommand{\geffds}{G_{\rm eff,dS}}
 
\newcommand{\ls}{\mathrel{\raise0.27ex\hbox{$<$}\kern-0.70em \lower0.71ex\hbox{{
$\scriptstyle \sim$}}}}
\newcommand{\ltgt}{\mathrel{\raise0.37ex\hbox{$<$}\kern-0.70em \lower0.71ex\hbox{{$>$}}}}
\newcommand{\gtlt}{\mathrel{\raise0.27ex\hbox{$>$}\kern-0.70em \lower0.71ex\hbox{{$<$}}}}

\newcommand{\n}{\nabla}

\begin{document}

\title{The Paths of Gravity in Galileon Cosmology} 
\author{Stephen Appleby$^1$ and Eric V.\ Linder$^{1,2}$} 
\affiliation{$^1$ Institute for the Early Universe WCU, Ewha Womans 
University, Seoul, Korea} 
\affiliation{$^2$ Berkeley Lab \& University of California, Berkeley, 
CA 94720, USA}

\begin{abstract}
Galileon gravity offers a robust gravitational theory for explaining 
cosmic acceleration, having a rich phenomenology of testable behaviors. 
We explore three classes of Galileon models -- standard uncoupled, and 
linearly or derivatively coupled to matter -- investigating the expansion 
history with particular attention to early time and late time attractors, 
as well as the linear perturbations.  From the relativistic and 
nonrelativistic Poisson equations we calculate the generalizations of 
the gravitational strength (Newton's constant), deriving its early 
and late time behavior.  By scanning through the parameters we derive 
distributions of the gravitational strength at various epochs and trace 
the paths of gravity in its evolution.  Using ghost-free and stability 
criteria we restrict the allowed parameter space, finding in particular 
that the linear and derivative coupled models are severely constrained by classical 
instabilities in the early universe.
\end{abstract}

\date{\today} 

\maketitle

\section{Introduction} 

The origin of the current cosmic acceleration is a fundamental mystery, 
even to whether it arises from a new field or a change in the laws of 
gravity.  Scalar-tensor gravity is an intriguing possibility, partaking 
of elements of both classes of explanation.  Those theories that do not 
have a scalar potential avoid some naturalness problems 
associated with this, and those which involve a symmetry or geometric 
origin moreover sidestep difficulties with high energy physics corrections. 

A class of theories possessing all these desiderata is Galileon 
gravity \cite{arXiv:0811.2197, arXiv:0901.1314,arXiv:0906.1967,arXiv:1007.5278}, 
involving a shift symmetry in the field, while the field itself can be 
viewed as a geometric object arising from higher dimensions and entering in 
invariant combinations that assure second order field equations and 
protect against various pathologies.  The derivative self couplings of 
the field also screen deviations from general relativity in high gradient 
regions (e.g.\ small scales or high densities), thus satisfying solar 
system and early universe constraints \cite{75796,hep-th/0106001,arXiv:1111.5090}. The 
galileon field, and closely related models \cite{arXiv:1106.2000,arXiv:1005.0868,arXiv:1006.0281,arXiv:1009.6151,arXiv:1008.0048,Qiu:2011cy,arXiv:1109.1047,arXiv:1103.5360,arXiv:1108.4242}, 
have been extensively studied in both theoretical \cite{arXiv:1008.1305,arXiv:1008.4128,arXiv:1008.4580,arXiv:1007.5424,arXiv:1008.3312,arXiv:1008.0745,arXiv:1003.5917,arXiv:1007.0443,arXiv:1104.0155} and 
observational \cite{arXiv:0909.4538,arXiv:1009.0677,arXiv:1003.3281,arXiv:1008.4236,arXiv:1011.6132,arXiv:1010.0407,arXiv:1007.2700,arXiv:1008.1588} contexts (and also with respect to inflation; \cite{Kobayashi:2010cm,arXiv:1009.2497,Kobayashi:2011nu}).

Galileon gravity thus merits investigation in detail, from both the 
expansion properties (homogeneous background) and structure growth 
(inhomogeneous perturbations) perspectives.  Regarding the expansion, we 
will particularly be interested in the early universe corrections to 
radiation and matter domination, the effective equation of state evolution, 
and late time attractor behavior, especially to a de Sitter state.  For 
the behavior of linearized perturbations (concentrating on subhorizon 
scales), we explore the form of the modified gravitational strengths 
(Newton constants) in the generalized Poisson equations, looking for 
characteristic features in their evolution. 

The parameter space of the theory will be constrained by general physical 
considerations, such as  positivity of the effective energy density and 
freedom from ghosts (Hamiltonian bounded from below).  Observationally, 
we ask that the radiation and matter eras remain intact and the effective 
dark energy density is of order 3/4 the critical density today.  These 
conditions still permit considerable diversity in the gravitational 
behavior, which is quite interesting to explore. 

In Section~\ref{sec:model} we briefly review the standard Galileon model 
and include terms with linear and derivative coupling to matter, setting 
up the field equations and modified Friedmann equations for a homogeneous, 
isotropic universe.  Section~\ref{sec:pert} sets up the linear perturbation 
theory and the generalized Poisson equations, deriving expressions for 
the gravitational strength modifications.  Solutions for the expansion 
history for the uncoupled, linear-, and derivative-coupled models are 
presented in Section~\ref{sec:expan}, including discussion of early and 
late time attractors.  Evolution of the gravitational strengths is studied 
in Section~\ref{sec:geff}, along with the no-ghost and stability conditions.  
We summarize the results and conclude in Section~\ref{sec:concl}.

\section{Classes of Galileon Gravity} \label{sec:model} 

Galileon gravity is a scalar field theory containing nonlinear derivative 
self couplings.  The action for the scalar field $\pi$ is invariant under 
Galilean symmetries $\pi \to \pi + c + b_{\mu}x^{\mu}$ in the absence of 
gravity, where $c$ and $b_{\mu}$ are a constant scalar and four vector 
respectively. The four dimensional action that preserves these symmetries 
contains five unique terms, consisting of scalar combinations of 
$\partial_{\mu} \pi$, $\partial_{\mu}\partial_{\nu} \pi$ and $\Box \pi$.  

These models share many similarities with the decoupling limit of DGP 
gravity \cite {hep-th/0005016,hep-th/0010186,astro-ph/0105068}
  and exhibit Vainshtein screening that restores the 
theory to the 
general relativity behavior under certain conditions.  Galileon gravity 
has field equations that contain at most second derivatives of the scalar 
field, and hence can be free of ghost instabilities. In addition, since 
these second derivative terms appear linearly, the Cauchy problem can be 
well defined.  

When gravity is introduced one must necessarily break the Galilean 
$\partial_{\mu} \pi \to \partial_{\mu} \pi + b_{\mu}$ symmetry.  However, 
a covariantized formulation of the Galileon model has been constructed in 
\cite{arXiv:0901.1314}, in which the action preserves the shift symmetry 
$\pi \to \pi + c$, and the Galilean symmetry is softly broken. 

The covariant Galileon action can be written as 
\begin{equation}\label{eq:m1}  
S = \int d^{4}x\,\sqrt{-g} \left[ { M_{\rm pl}^{2} R \over 2} - {1 \over 2}\sum_{i=1}^{5} c_{i}{\cal L}_{i}   - {\cal L}_{\rm m} \right]  
\end{equation}
where $c_{1-5}$ are arbitrary dimensionless constants, $g$ is the 
determinant of the metric, $M_{\rm pl}$ is the Planck mass, and $R$ is 
the Ricci scalar.  The Galileon Lagrangians are given by
\begin{eqnarray} & & {\cal L}_{1} = M^{3} \pi , \qquad {\cal L}_{2} = (\nabla_{\mu}\pi)(\nabla^{\mu}\pi), \qquad {\cal L}_{3} = (\Box \pi) (\nabla_{\mu}\pi)(\nabla^{\mu}\pi)/M^{3} \\ & & {\cal L}_{4} = (\nabla_{\mu}\pi)(\nabla^{\mu}\pi)\left[ 2(\Box\pi)^{2} - 2 \pi_{;\mu\nu}\pi^{;\mu\nu} - R (\nabla_{\mu}\pi)(\nabla^{\mu}\pi)/2 \right]/M^{6} \\ & & {\cal L}_{5} = (\nabla_{\mu}\pi)(\nabla^{\mu}\pi)\left[ (\Box\pi)^{3} - 3(\Box \pi)\pi_{;\mu\nu}\pi^{;\mu\nu} + 2 \pi_{;\mu}{}^{;\nu}\pi_{;\nu}{}^{;\rho}\pi_{;\rho}{}^{;\mu} - 6\pi_{;\mu}\pi^{;\mu\nu}\pi^{;\rho}G_{\nu\rho} \right] /M^{9} \end{eqnarray} 
where $M$ is the mass dimension taken out of the couplings to make the $c$'s 
dimensionless; without loss of generality 
we can take $M^{3} = M_{\rm pl}H_{0}^{2}$. 

The Galileon invariance severely restricts the form of the action, although 
further freedom remains in the coupling between $\pi$ and matter.  For 
example, the weak field limit of higher dimensional brane models typically 
involve a coupling of the form $\pi T$, where $T$ is the trace of the 
energy-momentum tensor; such a term explicitly enters into the effective 
scalar field action in the decoupling limit of the DGP model 
\cite{hep-th/0603199}.  We also 
consider a derivative coupling 
$T^{\mu\nu}\partial_{\mu}\pi \partial_{\nu}\pi$, which has other interesting 
properties and origins.  We will set the tadpole term to zero, i.e.\ 
$c_{1} = 0$, since we are interested in the effect of derivative self 
couplings on the growth and expansion histories and wish to avoid an 
explicit cosmological constant.

Thus, the action can be written as 
\begin{equation} \label{eq:ii1}   
S = \int d^{4}x\,\sqrt{-g} \left[   {M_{\rm pl}^{2} R \over 2} - {c_{2} \over 2} (\partial \pi)^{2} - {c_{3} \over M^{3}}(\partial \pi)^{2} \Box \pi - {c_{4}{\cal L}_{4} \over 2} - {c_{5}{\cal L}_{5} \over 2}  - {\cal L}_{\rm m} - {c_{\rm G} \over M_{\rm pl}M^{3}}T^{\mu\nu} \partial_{\mu}\pi \partial_{\nu} \pi - {c_{0} \over M_{\rm pl}} \pi T  \right] \,. 
\end{equation} 
For computational purposes we find it is easier to work in the Jordan frame, 
where the explicit coupling between $\pi$ and matter is removed via a metric 
redefinition.  In Appendix~\ref{sec:appT} we perform these metric 
transformations in the weak field limit, absorb like terms by renormalizing 
the $c$'s, and promote the final actions to their full, non-linear 
counterparts.  Here we state the result: 
\begin{equation} \label{eq:ii3}   
S = \int d^4 x\,\sqrt{-g} \left[ \left(1-2c_0\frac{\pi}{\mpl}\right) {M_{\rm pl}^{2} R \over 2} - {c_{2} \over 2} (\partial \pi)^{2} - {c_{3} \over M^{3}}(\partial \pi)^{2} \Box \pi - {c_{4}{\cal L}_{4} \over 2} - {c_{5}{\cal L}_{5} \over 2}   - {M_{\rm pl}\over M^{3}} c_{\rm G} G^{\mu\nu} \partial_{\mu}\pi\partial_{\nu}\pi  - {\cal L}_{\rm m}    \right] \,. 
\end{equation}

Varying the action with respect to the fields $g_{\mu\nu}$ and $\pi$, we 
obtain the general field equations that are exhibited in 
Appendix~\ref{sec:appfdeq}. 

In what follows we analyse three classes of models: the standard uncoupled 
Galileon, where $c_0=c_G=0$, the linearly coupled Galileon, where $c_G=0$, 
and the derivative coupled Galileon, where $c_0=0$.  Note that derivative 
coupling resulting in such $G^{\mu\nu} \partial_{\mu}\pi\partial_{\nu}\pi$ 
terms has been found of interest for dark energy in 
\cite{gubitosi,derham0610,acoleyen} and inflation in \cite{arXiv:1108.1406}, and has ties to higher dimensional, 
vector, and disformal gravity theories. 

To derive the cosmological evolution of these classes of Galileon gravity, 
we use the Robertson-Walker metric for a homogeneous, isotropic spacetime, 
\begin{equation} 
ds^{2} = -dt^{2} + a^{2} \delta_{ij}dx^{i}dx^{j} \,, 
\end{equation} 
where $a$ is the cosmic scale factor and we assume spatial flatness for 
simplicity.  For the homogeneous evolution $\pi = \pi(a)$ and there are 
two independent field equations.  

We can write the $(i,i)$ 
Einstein equation and $\pi$ dynamical equation in terms of dimensionless 
variables $y = \pi/M_{\rm pl}$, $\bar{H} = H/H_{0}$, and 
$x = \pi'/M_{\rm pl}$, where the Hubble parameter $H=\dot a/a$ and 
a prime denotes $d/d\ln a$, as a set of dynamical equations 
\begin{eqnarray} 
\label{eq:n8} x' &=& -x + {\alpha \lambda - \sigma \gamma \over \sigma\beta - \alpha\omega} \\ 
\label{eq:n9} \bar{H}' &=& - {\lambda \over \sigma} + {\omega \over \sigma} \left( { \sigma\gamma - \alpha\lambda \over \sigma\beta - \alpha\omega}\right) \\ 
\label{eq:n10} y' &=& x 
\end{eqnarray} 
where
\begin{eqnarray} 
& & \alpha = -3c_{3}\bar{H}^{3}x^{2}  + 15 c_{4} \bar{H}^{5} x^{3} + c_{0}\bar{H}  + {c_{2} \bar{H} x \over 6}  - {35 \over 2} c_{5} \bar{H}^{7} x^{4} -3c_{\rm G} \bar{H}^{3}x   \\ & & \gamma = 2c_{0} \bar{H}^{2}   - c_{3}\bar{H}^{4} x^{2}  + {c_{2} \bar{H}^{2} x \over 3 }  + {5 \over 2} c_{5} \bar{H}^{8} x^{4} - 2c_{\rm G}\bar{H}^{4}x  \\ & & \beta = -2 c_{3}\bar{H}^{4} x  + {c_{2} \bar{H}^{2} \over  6}  + 9 c_{4}\bar{H}^{6}x^{2} - 10c_{5} \bar{H}^{8} x^{3} - c_{\rm G} \bar{H}^{4} \\ 
& & \sigma = 2 ( 1 - 2c_{0}y)\bar{H} -2c_0 \bar{H} x + 2c_{3}\bar{H}^{3} x^{3} 
-15c_{4} \bar{H}^{5}x^{4} + 21c_{5} \bar{H}^{7} x^{5} +6c_{\rm G}\bar{H}^{3}x^{2} \\ \nonumber & &  \lambda = 3( 1 - 2c_{0}y) \bar{H}^{2}  - 2 c_{0} \bar{H}^{2}x   \\ & & \qquad \qquad  - 2c_{3} \bar{H}^{4} x^{3}  + {c_{2} \bar{H}^{2} x^{2} \over 2}+ {\Omega_{\rm r0} \over a^{4}} + {15 \over 2} c_{4} \bar{H}^{6}x^{4} -9c_{5} \bar{H}^{8}x^{5} -c_{\rm G}\bar{H}^{4}x^{2} \\ 
& & \omega = -2c_0 \bar{H}^2 + 2c_{3} \bar{H}^{4}x^{2} 
- 12c_{4}\bar{H}^{6}x^{3} + 15c_{5} \bar{H}^{8}x^{4} + 4c_{\rm G}\bar{H}^{4}x   \ . 
\end{eqnarray} 

As a check of our numerical solutions we verify that the redundant 
Friedmann equation is satisfied at all times during the evolution: 
\begin{equation} 
\left( 1 - 2c_{0}y\right) \bar{H}^{2} = {\Omega_{m 0 } \over a^{3}}  + {\Omega_{\rm r0} \over a^{4}} + 2c_{0} \bar{H}^{2} x + {c_{2} \over 6} \bar{H}^{2} x^{2} - 2 c_{3} \bar{H}^{4} x^{3} + {15 \over 2} c_{4} \bar{H}^{6} x^{4}  - 7 c_{5} \bar{H}^{8}x^{5} - 3c_{\rm G} \bar{H}^{4} x^{2} \,, 
\label{eq:fried} 
\end{equation} 
where $\Omega_{m0}$ and $\Omega_{r0}$ are the present matter and radiation 
energy densities, respectively, in units of the critical density. 

It is also useful to write the energy density and pressure of the scalar 
field: 
\begin{eqnarray} 
\label{eq:o1} & & {\rho_{\pi} \over H_{0}^{2}M_{\rm pl}^{2}} =  6c_{0}\bar{H}^{2} x + {c_{2} \over 2} \bar{H}^{2} x^{2} - 6 c_{3} \bar{H}^{4} x^{3}   + {45 \over 2} c_{4} \bar{H}^{6} x^{4} - 21c_{5} \bar{H}^{8}x^{5} - 9c_{\rm G} \bar{H}^{4} x^{2}  \\ 
\nonumber & &  {P_{\pi} \over H_{0}^{2}M_{\rm pl}^{2}} = -c_{0} \left[ 4\bar{H}^{2} x  + 2\bar{H} \left( \bar{H}  x \right)' \right] + {c_{2} \over 2} \bar{H}^{2} x^{2} + 2c_{3} \bar{H}^{3} x^{2} \left( \bar{H}  x \right)' - c_{4} \left[ {9 \over 2} \bar{H}^{6}x^{4} + 12 \bar{H}^{6}x^{3}x' + 15 \bar{H}^{5} x^{4} \bar{H}' \right]  \\  \label{eq:o2} & & \hspace{40mm}  + 3c_{5} \bar{H}^{7} x^{4} \left( 5 \bar{H} x' + 7 \bar{H}' x + 2\bar{H} x \right) + c_{\rm G} \left[ 6\bar{H}^{3}x^{2}\bar{H}' + 4\bar{H}^{4}x x' + 3 \bar{H}^{4} x^{2} \right]   
\end{eqnarray} 
from which one can define an effective equation of state parameter 
for the Galileon: $w \equiv P_{\pi}/\rho_{\pi}$.

\section{Inhomogeneous Perturbations and Modified Poisson Equations} \label{sec:pert} 

While the previous section set up the background dynamics for the 
cosmological evolution, we also need the inhomogeneous equations to 
investigate the effects of perturbations on the cosmic growth history 
and gravitational modifications to the generalized Poisson equations. 
We here consider linear perturbation theory for scalar modes in the 
subhorizon limit.  In the Newtonian gauge the perturbed metric is 
\begin{equation} 
ds^{2} = -(1+2\psi) dt^{2} + a^{2} (1-2\phi) \delta_{ij}dx^{i}dx^{j} \,. 
\end{equation} 
 
The linearized Einstein and scalar field equations are exhibited in 
Appendix~\ref{sec:applineq}.  These lead to the modified Poisson equations 
\begin{eqnarray} 
& & \bar{\nabla}^{2} \phi = {4\pi a^{2} G^{(\phi)}_{\rm eff} \rho_{\rm m} \over H_{0}^{2}}\delta_{\rm m}  \\  & &   \bar{\nabla}^{2} \psi = {4\pi a^{2} G^{(\psi)}_{\rm eff} \rho_{\rm m} \over H_{0}^{2}}\delta_{\rm m} \\  & &   \bar{\nabla}^{2} (\psi+\phi) = {8\pi a^{2} G^{(\psi+\phi)}_{\rm eff} \rho_{\rm m} \over H_{0}^{2}}\delta_{\rm m}  \,, 
\end{eqnarray} 
where we describe the effect of the Galileon on subhorizon density 
perturbations through modifications $\geff$ of the gravitational strength, or 
effective Newton's constant.  These will be functions of the background 
quantities $\bar{H}$, $y$, and their derivatives. 

Note that $\geff^{(\psi)}$ is central to the growth of cosmic structure 
and is called ${\mathcal V}$ in \cite{daniellinder} and $\mu$ in \cite{zhao}, 
while $\geff^{(\psi+\phi)}$ is important for light deflection and the 
integrated Sachs-Wolfe effect, and called ${\mathcal G}$ and $\Sigma$ in 
those two references, respectively.  Translation tables between 
other parametrizations are given in \cite{roysoc,daniellinder}.  One can 
regard the $\psi$ Poisson equation as governing nonrelativistic geodesics 
and the $\psi+\phi$ Poisson equation as governing relativistic geodesics 
in the inhomogeneous spacetime \cite{bertschinger11114659,Dossett:2011zp}. 

The modified gravitational strengths are given in terms of Newton's 
constant $G_N$ by 
\begin{eqnarray} 
\label{eq:geffc4}  \geff^{(\phi)} &=& { 2 \left( \kappa_{4}\kappa_{6} - \kappa_{5}\kappa_{1}\right)  \over \kappa_{5} \left( \kappa_{4} \kappa_{1} - \kappa_{5} \kappa_{3} \right) - \kappa_{4} \left( \kappa_{4}\kappa_{6} - \kappa_{5}\kappa_{1}\right) }\ G_{\rm N} \\ 
\label{eq:geffc5}  \geff^{(\psi)} &=& { 4 \left( \kappa_{3}\kappa_{6} - \kappa_{1}^{2}\right)  \over \kappa_{5} \left( \kappa_{4} \kappa_{1} - \kappa_{5} \kappa_{3} \right) - \kappa_{4} \left( \kappa_{4}\kappa_{6} - \kappa_{5}\kappa_{1}\right) }\ G_{\rm N} \\ 
 \geff^{(\psi+\phi)} &=& {  \kappa_{6} \left(2\kappa_{3} + \kappa_{4}\right) - \kappa_{1}\left( 2\kappa_{1} + \kappa_{5}\right)  \over \kappa_{5} \left( \kappa_{4} \kappa_{1} - \kappa_{5} \kappa_{3} \right) - \kappa_{4} \left( \kappa_{4}\kappa_{6} - \kappa_{5}\kappa_{1}\right) }\ G_{\rm N}
\end{eqnarray} 
where $\kappa_{1-6}$ are functions of the background variables: 
\begin{eqnarray} 
\nonumber & & \kappa_{1} = -6 c_{4} \bar{H}^{3}x^{2} \left( \bar{H}' x + \bar{H} x' + {\bar{H} x \over 3} \right) + 2c_{\rm G} \left( \bar{H}\bar{H}' x + \bar{H}^{2} x' + \bar{H}^{2}x \right) - 2c_{0} +c_{5} \bar{H}^{5} x^{3} \left( 12 \bar{H} x'  + 15 \bar{H}' x + 3\bar{H} x \right)   \\ \nonumber & & \kappa_{2} = -{c_{2} \over 2} + 6c_{3}\bar{H}^{2} x + 3 c_{\rm G} \bar{H}^{2} - 27c_{4} \bar{H}^{4}x^{2} + 30 c_{5} \bar{H}^{6} x^{3}
  \\ \nonumber & & \kappa_{3} = -( 1 -2c_{0}y) - {c_{4} \over 2} \bar{H}^{4} x^{4} + c_{\rm G}\bar{H}^{2}x^{2} - 3c_{5} \bar{H}^{5}x^{4}\left(\bar{H}x' + \bar{H}'x \right)  \\ \nonumber & & \kappa_{4} = - 2( 1 - 2c_{0}y) + 3c_{4}\bar{H}^{4} x^{4}  -2c_{\rm G}\bar{H}^{2}x^{2} -6c_{5} \bar{H}^{6}x^{5} \\ 
\nonumber & & \kappa_{5} = 2c_{3} \bar{H}^{2} x^{2} - 12c_{4} \bar{H}^{4} x^{3}  + 4c_{\rm G}\bar{H}^{2} x -2c_{0} + 15c_{5}\bar{H}^{6}x^{4} \\ 
\nonumber & & \kappa_{6} =  {c_{2} \over 2} - 2c_{3}  \left( \bar{H}^{2} x' + \bar{H}\bar{H}' x + 2\bar{H}^{2}x \right) + c_{4}  \left( 12\bar{H}^{4}x x' + 18\bar{H}^{3}x^{2}\bar{H}'  + 13 \bar{H}^{4}x^{2} \right)  \\ 
\nonumber & &\hspace{10mm} - c_{\rm G} \left( 2\bar{H}\bar{H}' + 3\bar{H}^{2} \right) - c_{5}  \left( 18\bar{H}^{6}x^{2}x' + 30\bar{H}^{5}x^{3}\bar{H}' + 12\bar{H}^{6}x^{3} \right)  \ . 
\end{eqnarray} 

The general relativistic form of the Poisson equation is recovered if we 
set $c_{0} = c_{3-5} = c_{\rm G} = 0$ ($c_{2}$ can remain unspecified; in 
this limit we have a massless, minimally coupled scalar field which will 
not modify the effective Newton constants).  In that case, 
$\geff^{(\psi)} = \geff^{(\phi)} = \geff^{(\psi+\phi)} = G_{\rm N}$.

\section{Expansion History in Galileon Cosmology} \label{sec:expan}  

We now evaluate how Galileon cosmology affects the cosmic expansion 
history relative to the standard $\Lambda$CDM paradigm, including early 
universe asymptotes, the onset of cosmic acceleration, and late time 
attractors, especially to a de Sitter state.

\subsection{\label{sec:ub}Uncoupled Galileon: $c_0=0=c_G$} 

The first, simplest case corresponds to the absence of any direct coupling 
between the scalar field $\pi$ and matter; the parameters are then 
$c_2,c_3,c_4,c_5$.  Such a model has been extensively studied in 
\cite{arXiv:1010.0407,arXiv:1011.6132,arXiv:1007.2700}.  Equations~(\ref{eq:n8})-(\ref{eq:n10}) 
describing the 
background evolution define an autonomous system, and it is straightforward 
to establish the existence of asymptotic fixed points. 

The only fixed points for matter and radiation domination correspond to 
$(\Omega_m,\Omega_r,x)=(1,0,0)$ and $(0,1,0)$ respectively.  We therefore 
solve the $\pi$ equation assuming matter and radiation domination to obtain 
the behaviour of $\rho_{\pi}$ at early times.  For late times, we want 
cosmic acceleration and so we look for a late time de Sitter asymptotic 
state by finding the fixed points of the dynamical system $(\bar{H},x)$, 
taking $\Omega_m = \Omega_r = 0$. 

\begin{figure}
\centering
\mbox{\resizebox{0.49\textwidth}{!}{\includegraphics[angle=270]{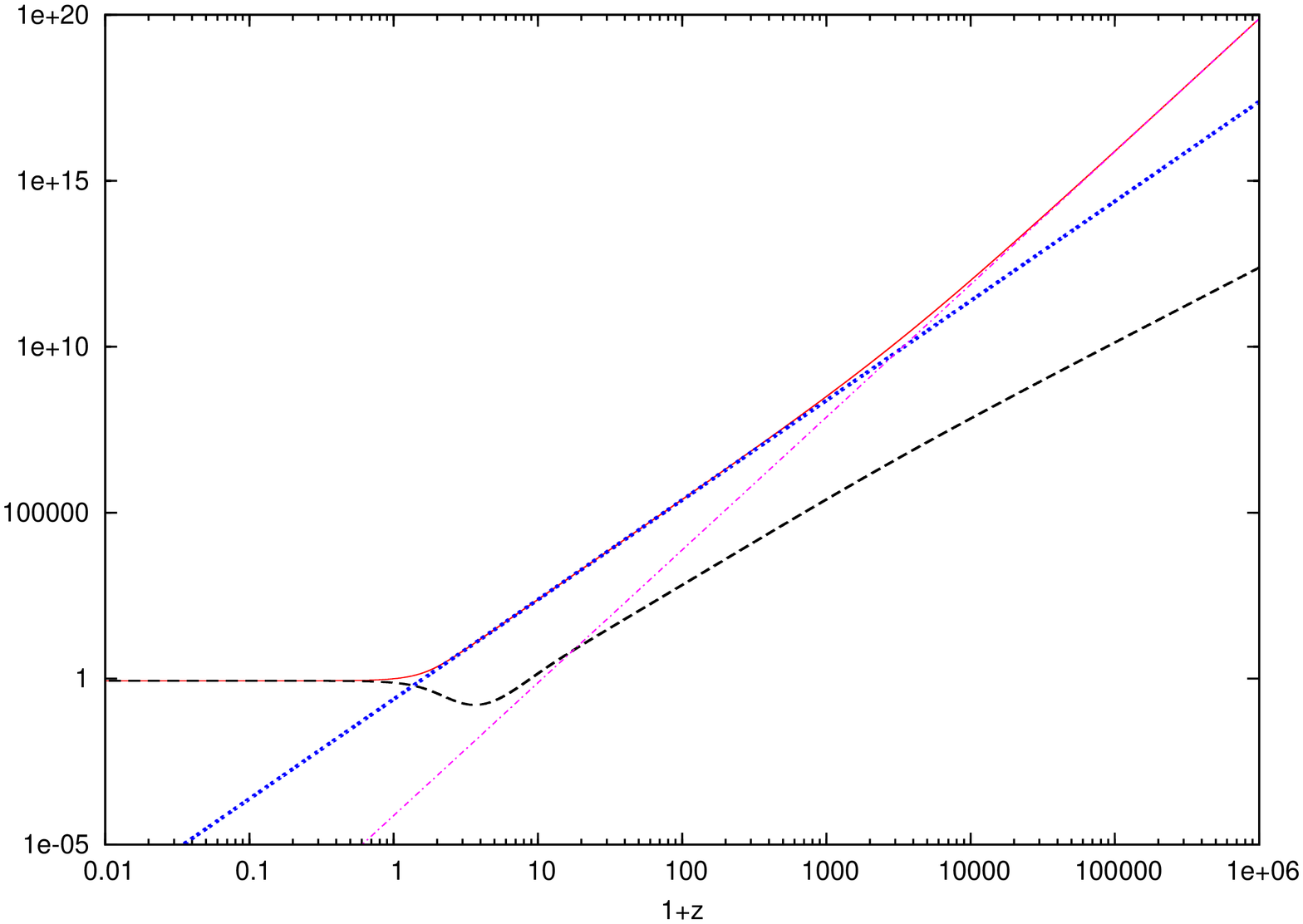}}}
\mbox{\resizebox{0.49\textwidth}{!}{\includegraphics[angle=270]{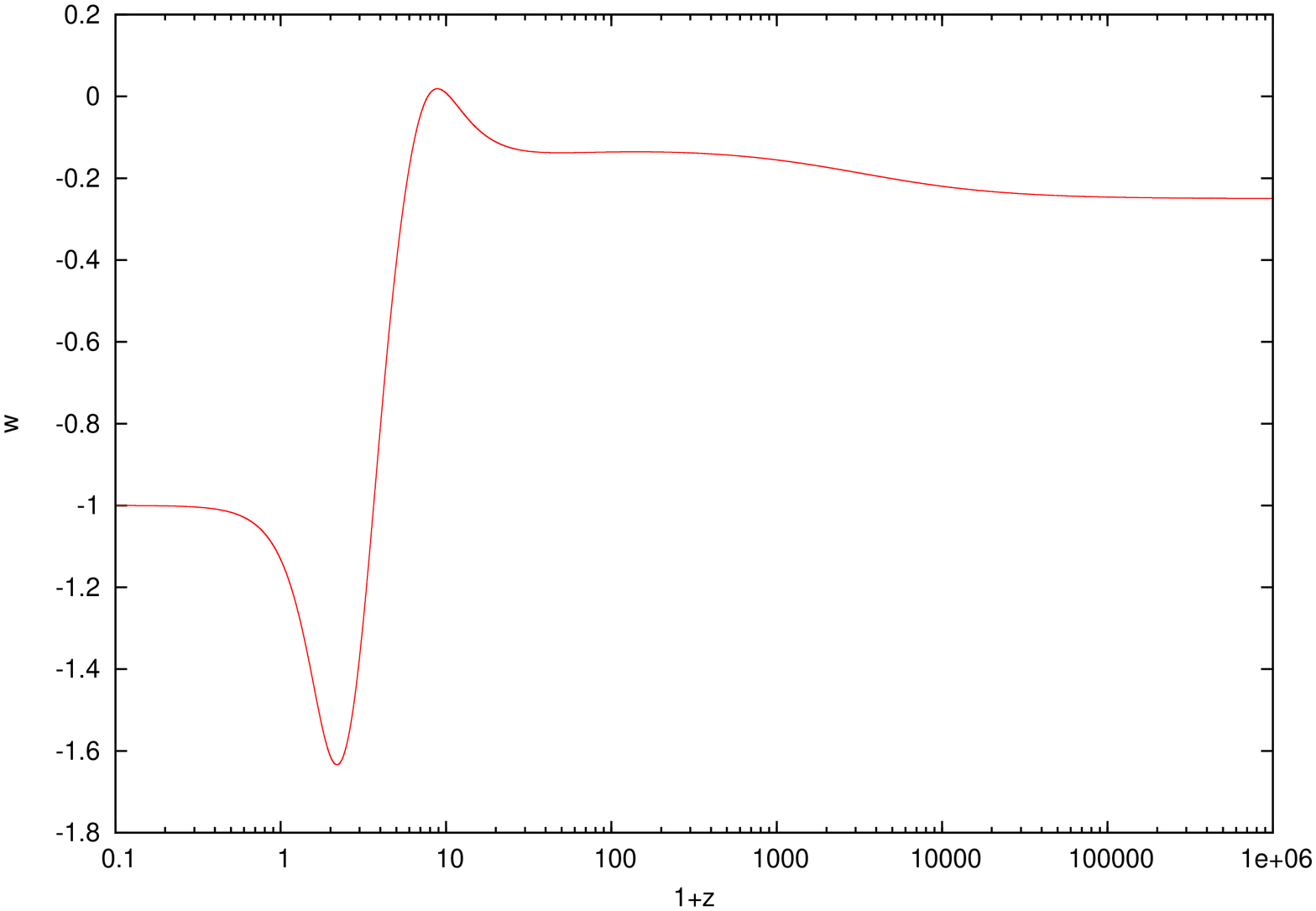}}}
\caption{ [Left panel] The redshift evolution of $\bar{H}^{2}$ 
(solid red), $8\pi G \rho_{m}/3H_{0}^{2}$ (dotted blue), 
$8\pi G \rho_r/3H_{0}^{2}$ (dot-dashed magenta) and 
$8\pi G \rho_{\pi}/3H_{0}^{2}$ (long dash black) for the uncoupled 
Galileon, with parameters $c_{2} = -27.3$, $c_{3}=-12.8$, $c_{4} = -1.7$, 
$c_{5} = -1$.  Here, the scalar field energy density slowly grows relative 
to matter at early times, as expected from the analytic results. The $c_{5}$ 
term dominates the $\pi$ dynamics up to $z\sim 10$, after which we observe 
a turnaround to approach the late time de Sitter point. [Right panel] The 
effective equation of state of the Galileon field.  During radiation 
domination $w=-1/4$, in agreement with the analytic results; the model 
currently accelerates the expansion and approaches a late time de Sitter 
point.} 
\label{fig:expunc} 
\end{figure}

The full numerical solutions are illustrated in Figure~\ref{fig:expunc} 
for specific parameter choices, 
and agree asymptotically with the fixed point behaviors.  We plot 
the time evolution of $\bar{H}^2$, $\rho_m/(3H_{0}^{2}M_{\rm pl}^{2})$, 
$\rho_r/(3H_{0}^{2}M_{\rm pl}^{2})$ and $\rho_{\pi}/(3H_{0}^2\mpl^2)$ in 
the left panel, and the effective dark energy equation of state parameter 
$w$ in the 
right panel.

To calculate the behaviour of the $\pi$ field at early times we look for 
solutions taking $\bar{H} \sim a^{-2}$ (during radiation domination) or 
$\bar{H} \sim a^{-3/2}$ (during matter domination).  The field equation 
of motion (\ref{eq:n8}) becomes 
\begin{equation} 
\label{eq:nm1} x' = -x +  { -5c_{3}\bar{H}^{4}x^{2} + 30 c_{4}\bar{H}^{6}x^{3} - 75c_{5}\bar{H}^{8}x^{4}/2  \over c_{2} \bar{H}^{2}/6 - 2c_{3}\bar{H}^{4}x + 9 c_{4} \bar{H}^{6} x^{2} - 10c_{5}\bar{H}^{8}x^{3}} 
\end{equation} 
during radiation domination and 
\begin{equation} 
\label{eq:nm2} x' = -x + {-c_{2} \bar{H}^{2}x/12 - 7c_{3}\bar{H}^{4}x^{2}/2 + 45c_{4}\bar{H}^{6}x^{3}/2  - 115c_{5}\bar{H}^{8}x^{4}/4 \over c_{2}\bar{H}^{2}/6 - 2c_{3}\bar{H}^{4}x + 9c_{4}\bar{H}^{6}x^{2} -10c_{5}\bar{H}^{8}x^{3} } 
\end{equation} 
during matter domination.  We note the existence of a solution 
$x \sim \bar{H}^{-2}$ during both radiation and matter domination; this 
behaviour was first observed in \cite{arXiv:1007.2700}.  However, solutions close to 
this fixed point at early times require an extraordinary fine tuning of 
initial conditions to ensure that the $\pi$ field energy density remains 
subdominant until late times, as the energy density of the $\pi$ field 
grows at a fast rate relative to matter and radiation (the $\pi$ field 
equation of state is $w= -7/3$ and $w= -2$ during 
radiation and matter domination respectively.)  We are interested in the 
behaviour of the $\pi$ field for generic initial conditions. 

Since each $c_{2-5}$ contribution to the scalar field energy density 
enters in the ratio of $\bar{H}^{2} x$ relative to the previous one, 
whether the initial condition has $x_{\rm i} \ll \bar{H}_{\rm i}^{-2}$ 
or not determines the early time evolution.  For example, if this holds 
then the lowest, $c_{2}$ term will dominate $\rho_{\pi}$ and $x\sim a^{-1}$, 
$x \sim a^{-3/2}$ during radiation and matter domination respectively.  
Since the other terms then decay at a faster rate than the $c_{2}$ term, 
once $c_{2}$ dominates it will do so at all subsequent times.  
In this case the $\pi$ energy density will decay as a massless scalar field 
$\rho_{\pi} \sim a^{-6}$ at all times (recall $c_2$ involves the canonical 
kinetic term), leading to no interesting dynamics. 

As we increase $x_{\rm i}$ the $c_{3-5}$ terms will dominate. This is 
expected in the early Universe, 
where the model will be in the strongly (self)coupled regime. Note that 
the dynamics of $\rho_{\pi}$ is hierarchical: for example if the $c_{3}$ 
term dominates at a particular time then the $c_{4,5}$ terms will remain 
subdominant at all subsequent times, however the $c_{2}$ term grows relative 
to the $c_{3}$ term and can eventually dominate. 

During the radiation epoch, we find the following behaviour of $x$ and 
$\rho_{\pi}$, depending on which of the $c_{2-5}$ terms dominate: 
\begin{eqnarray} 
\label{eq:i10} c_{5} :&& \quad x \sim a^{11/4} \qquad \rho_{\pi} \sim 
c_{5} a^{-9/4} \qquad w=-1/4  \\ 
\label{eq:i1} c_{4} :&& \quad x \sim a^{7/3} \qquad \rho_{\pi} \sim 
c_{4} a^{-8/3} \qquad w=-1/9 \\ 
\label{eq:i2} c_{3} :&& \quad x \sim a^{3/2} \qquad \rho_{\pi} \sim 
c_{3} a^{-7/2} \qquad w=1/6 \\ 
\label{eq:i3} c_{2} :&& \quad x \sim a^{-1} \qquad \rho_{\pi} \sim  
c_{2}a^{-6} \qquad\quad w=1 
\end{eqnarray}
and during the matter epoch 
\begin{eqnarray}  
\label{eq:i11} & & c_{5} : \quad x \sim a^{15/8} \qquad \rho_{\pi} \sim c_{5} a^{-21/8} \qquad w=-1/8 \\ 
\label{eq:i4} & & c_{4} : \quad x \sim a^{3/2} \qquad \rho_{\pi} \sim c_{4} a^{-3} \qquad\quad w=0 \\ 
\label{eq:i5}& & c_{3} : \quad x \sim a^{3/4} \qquad \rho_{\pi} \sim c_{3} a^{-15/4} \qquad w=1/4 \\ 
\label{eq:i6} & & c_{2} : \quad x \sim a^{-3/2} \qquad \rho_{\pi} \sim c_{2}a^{-6} \qquad\quad w=1 \,. 
\end{eqnarray} 

The effective equation of state of the $\pi$ field varies between 
$w \in[-1/4,1/4]$ when the $c_{3-5}$ terms dominate during these 
epochs. During radiation domination, the $\pi$ energy density typically 
grows relative to the matter component, however this effect is generally 
small and does not require a significant tuning of $\Omega_{\pi} (z_i)$ 
initially. For example, if the $c_{5}$ term provides the main contribution 
to $\rho_{\pi}$ during radiation domination then we have $\rho_{\pi}/\rho_m 
\sim a^{3/8}$. This is the worst case scenario, and provides a limit on how 
large we can set $\rho_{\pi}$ initially such that it remains subdominant at 
matter radiation equality (the exact constraint depends on our choice of 
$z_i$).  Thus Galileon models possess the interesting property that they 
can belong to the class of early dark energy models, in the sense that 
their energy density is non-negligible at early times for generic initial 
conditions.  (Indeed, the $c_4$ dominating case maintains a constant fraction 
of the matter density during matter domination.) 

To calculate the behaviour of the $\pi$ field at late times one can 
obtain the asymptotic de Sitter fixed points using a dynamical systems 
approach based on Eqs.~(\ref{eq:n8})-(\ref{eq:n9}).  The fixed points 
correspond to solutions of the algebraic relations 
\begin{eqnarray} 
\label{eq:na1} & & \lambda = -\omega x \\ 
\label{eq:na2} & & \gamma = -\beta x \end{eqnarray}
that are coupled polynomials for $\bar{H}^2$ and $x$.  For any values of 
the parameters $c_{2-5}$ there is a double zero corresponding to Minkowski 
space; $\bar{H} = 0$. In addition, one can show that there are at most three 
de Sitter, three anti de Sitter and six unphysical complex solutions for 
$\bar{H}$. To see this, we can use the coupled parameter $\chi=\bar H^2 x$, 
in which case Eq.~($\ref{eq:na2}$) reduces to the cubic polynomial  
\begin{equation} 
\label{eq:na6} 15c_{5} \chi^{3} -18 c_{4} \chi^{2} + 6 c_{3}\chi -c_{2}=0 \ , 
\end{equation} 
\noindent which has three (one) real solutions if its discriminant 
\begin{equation} 
D =  144c_{3}^{2}c_{4}^{2} - 160c_{3}^{3}c_{5} + 360c_{2}c_{3}c_{4}c_{5} - 288c_{2}c_{4}^{3} -75c_{2}^{2}c_{5}^{2}  
\end{equation}
is positive (negative). One can also state that for $c_{5} < 0$ (as needed 
for positive energy density when $c_5$ dominates early) 
there is exactly one real solution for $x$ 
whenever $5c_{3}c_{5} - 6c_{4}^{2} > 0$. Note that this does not 
necessarily imply the existence of a de Sitter point, which can only be 
determined by solving Eq.~(\ref{eq:na1}). In terms of $\chi$, the condition 
for the existence of a de Sitter vacuum state is 
\begin{equation} 
\label{eq:na5} 3\bar{H}^{4} = -6c_{5}\chi^{5} + {9 \over 2} c_{4}\chi^{4} 
- {c_{2} \over 2} \chi^{2} > 0 \ . 
\end{equation} 

Hence for any real $\chi$, there is at most one positive and one negative 
real $\bar{H}$ solution to Eq.~(\ref{eq:na5}), and there are at most 
three distinct, real $\chi$ solutions to Eq.~(\ref{eq:na6}). The minimal 
number of de Sitter solutions is zero.

Certain specific cases are worth mentioning. We require at least two 
non-zero $c_{2-5}$ terms to admit a de Sitter fixed point, and during the 
approach to the de Sitter asymptote, generically all terms are of the same 
order. Restricting our analysis to the case that any two of the terms 
$c_{2-5}$  are non-zero, we can succinctly write the conditions for 
positivity of $\rho_{\pi}$ at early times and the existence of a late time 
de Sitter point as: $(c_{2},c_{3}) = (-,-)$, $(c_{2},c_{4}) = (-,+)$, 
$(c_{2},c_{5}) = (-,-)$. In each case, there is exactly one de Sitter 
point (although in the $c_{4}$ case the solution has multiplicity two).  
Note that the standard kinetic term must exist (or the no-ghost condition 
of Sec.~\ref{sec:noghostmain} is violated) and must take the non-canonical 
sign. 

Both the early time and late time analytic results are borne out by 
the numerical solutions of the dynamical equations.  Throughout this 
article we take initial conditions at $z_i=10^6$ of $\rho_\pi(z_i)=10^{-5} 
\rho_m(z_i)$ and fix $\Omega_m(z=0)=0.24$.  The parameters, and behaviors, 
exhibited in Figure~\ref{fig:expunc} are typical in the sense elucidated 
in Section~\ref{sec:paths}.

\subsection{\label{sec:lin}Linearly Coupled Galileon: $c_0\ne 0$, $c_G=0$} 

We now consider a linear coupling between the field $\pi$ and the trace 
of the energy-momentum tensor (or Ricci scalar) \cite{arXiv:0905.1325,arXiv:0909.4538}.  As mentioned, such a 
scenario is typical in higher dimensional braneworld models, and in 
particular in the decoupling limit of the DGP model.  We begin as before 
by examining the behaviour of the $\pi$ field during matter and radiation 
domination. We initially consider the DGP like case where $c_{4,5}=0$ and 
then extend to include the more general Galileon kinetic terms.

In the decoupling limit, the DGP model can be written as a four dimensional 
effective action with a scalar field coupled to the Ricci scalar and 
containing kinetic terms of the form $c_{2,3}$. 
With the parameters $c_0,c_2,c_3$ we 
expect to recover DGP-like behaviour in the early universe, where 
$\rho_{\pi}$ is subdominant and we can neglect the backreaction of the 
$\pi$ field on the geometry. Indeed, if we choose initial conditions such 
that $c_{0}y \ll 1$, we find two approximate solutions during matter 
domination of the form $\bar{H} \sim a^{-3/2}$, $x = \pm A_{0}\bar{H}^{-1}$, 
where $A_{0} = \sqrt{c_{0}/(3c_{3})}$. On this solution, $\rho_{\pi} \sim 
\pm 4 c_{0} A_{0} \bar{H}$, the familiar modification to the Friedmann 
equation one would expect from a model mimicking the DGP model (here we 
choose the sign convention $c_{0}, c_{3} >  0$).  

On the ``self accelerating branch'' corresponding to $x > 0$, $\rho_\pi>0$, 
one can also see that the $\pi$ field does not satisfy the no-ghost 
condition given by (see Sec.~\ref{sec:noghostmain}) 
\begin{equation} 
\left( c_2 -12c_{3} \bar{H}^{2}x \right)(1 - 2c_{0} y) + 
6 \left( c_{3} \bar{H}^{2}x^{2} -c_0 \right)^{2} > 0  \,. 
\end{equation} 
At early times the first term on the left hand side dominates, and we can 
neglect the $c_{2}$ and $c_{0}y$ contributions, leaving just 
$-12c_3 \bar{H}^2 x>0$.  For positive $c_{3}$, the sign of $x$ dictates 
whether the field is a ghost; on the self accelerating branch the inequality 
is not satisfied and $\pi$ is a ghost as expected \cite{hep-th/0604086,hep-th/0303116}. 

We now introduce the $c_{4,5}$ terms, and consider how the evolution of 
$\pi$ is modified. To exhibit the behaviour of $x$ at early times, we 
write $\bar{H} \sim a^{-2}$ and $\bar{H} \sim a^{-3/2}$ for radiation 
and matter domination.  But it should be observed that the standard 
behaviour of $\bar{H}$ is no longer guaranteed: from the Friedmann equation 
(\ref{eq:fried}) it is clear that to obtain the standard behaviours 
$\bar{H} \sim a^{-2}$ and $\bar{H} \sim a^{-3/2}$ we must impose 
$c_0 y \ll 1$ to avoid modification of the left hand side.  We therefore 
set $y = 0$ initially, at some redshift deep in radiation domination, and 
check that $y$ remains small. 

During radiation domination, we find the same solutions 
(\ref{eq:i10})-(\ref{eq:i3}) as in the uncoupled case due to the fact 
that the $c_{0}$ coupling term is subdominant in the $x$ equation, so  
\begin{equation} 
\label{eq:nml1} x' = -x + { -5c_{3}\bar{H}^{4}x^{2} + 30 c_{4}\bar{H}^{6}x^{3} - 75c_{5}\bar{H}^{8}x^{4}/2  \over c_{2} \bar{H}^{2}/6 - 2c_{3}\bar{H}^{4}x + 9 c_{4} \bar{H}^{6} x^{2} - 10c_{5}\bar{H}^{8}x^{3}} 
\ \ . 
\end{equation} 
However during matter domination we find modified behaviour, 
\begin{equation} 
\label{eq:nm20}  x' = -x + {-c_{2} \bar{H}^{2}x/12 - 7c_{3}\bar{H}^{4}x^{2}/2 + 45c_{4}\bar{H}^{6}x^{3}/2 -c_{0} \bar{H}^{2} /2 - 115c_{5}\bar{H}^{8}x^{4}/4 \over c_{2}\bar{H}^{2}/6 - 2c_{3}\bar{H}^{4}x + 9c_{4}\bar{H}^{6}x^{2} -10c_{5}\bar{H}^{8}x^{3} } 
\ \ . 
\end{equation} 

The behaviour of $x$ and hence $\rho_{\pi}$ depends upon the initial 
conditions imposed, and general analytic solutions do not exist.  However 
for certain limiting cases one can construct explicit solutions.  One 
example is when the $c_{3}$ term dominates, corresponding to the DGP like 
solution discussed above with 
$\rho_{\pi}/(H_{0}^{2}M_{\rm pl}^{2}) = 4c_{0}A_{0}\bar{H}$. Analogous 
solutions exist for the cases where other terms dominate; note that 
again the contributions to $\rho_\pi$ are hierarchical -- if the $c_n$ 
term dominates early, it can only be superceded later in matter 
domination by a $c_m$ term with $0<m<n$.  The matter dominated era 
solutions, depending on 
which of the $c_{2-5}$ terms dominate alongside $c_0$, are 
\begin{eqnarray} 
\label{eq:l10} & & c_{5},c_{0} : \qquad x = A_{0} \bar{H}^{-3/2} \qquad \rho_{\pi} ={16 \over 5} c_{0}A_{0}\bar{H}^{1/2} \qquad A_{0} =  \left({2c_{0} \over 15c_{5}}\right)^{1/4} \qquad w=-{11 \over 8}  \\ 
\label{eq:llin1} & & c_{4},c_{0} : \qquad x = A_{0} \bar{H}^{-4/3} \qquad \rho_{\pi} ={7 \over 2} c_{0}A_{0}\bar{H}^{2/3} \qquad A_{0} = \left(-{c_{0} \over 9c_{4}}\right)^{1/3} \qquad w=- {26 \over 21} \\ 
\label{eq:l2} & & c_{3},c_{0} : \qquad x = A_{0} \bar{H}^{-1} \qquad \rho_{\pi} =4c_{0} A_{0}\bar{H} \qquad \qquad A_{0} =  \left(c_{0} \over 3c_{3}\right)^{1/2} \qquad\quad w=-1 \\ 
\label{eq:c20} & & c_2,c_0 : \qquad x=A_0 \qquad\quad \rho_\pi=5c_0 A_{0} \bar{H}^2 
\qquad\qquad\quad A_0=\frac{-2c_0}{c_2} \qquad\qquad \quad w=-{ 2 \over 5}  \,, 
\end{eqnarray} 
where we have chosen sign conventions such that $x > 0$.  Note that 
because of the explicit coupling to the Ricci scalar, the standard 
continuity equation and $\rho_\pi\sim a^{-3(1+w)}$ do not hold.  
We also note that 
the last case is only viable if $|c_0^2/c_2|\ll 1$, 
otherwise we will pick up non-negligible corrections to the standard 
matter era due to the presence of the $c_{0}y$ term on the left hand side 
of the Friedmann equation. 
We will show in Appendix~\ref{sec:appghost} that in each of the other cases, 
the $\pi$ field is a ghost whenever $\rho_{\pi} > 0$ (this result does 
not depend on our choice $x >0$).  Therefore the linear coupled model 
is only viable when either the $c_2$ term dominates the $c_{m>2}$ terms at 
early times and $|c_0^2/c_2|\ll 1$, or when the $c_{2-5}$ contributions to 
the $\pi$ energy density are much larger than that from $c_{0}$. In the latter 
scenario the model will evolve approximately as in the uncoupled case 
during matter and radiation domination. 

At late times we no longer obtain a de Sitter fixed point; this is due to 
the fact that the field $y$ now enters the Friedmann equation explicitly.  
The de Sitter solution in the uncoupled case corresponds to $\bar{H},x$ 
approaching constant values asymptotically, and this is no longer possible 
since then $c_0 y$ would grow.   However, since $y$ grows as $\ln a$ for 
constant $x$, a near de Sitter state is possible, where $w\sim -1$ 
for $z \lesssim -0.5$. 
From our numerical studies we find that typical deviations from 
$w = -1$ are less than $\sim 5\%$ for model parameters in the range 
$c_{2-5} \sim (-10,10)$ and $c_{0} \sim (-1,1)$ at $z \sim -0.95$ 
(though larger deviations are possible). 

\begin{figure}
\centering
\mbox{\resizebox{0.48\textwidth}{!}{\includegraphics[angle=270]{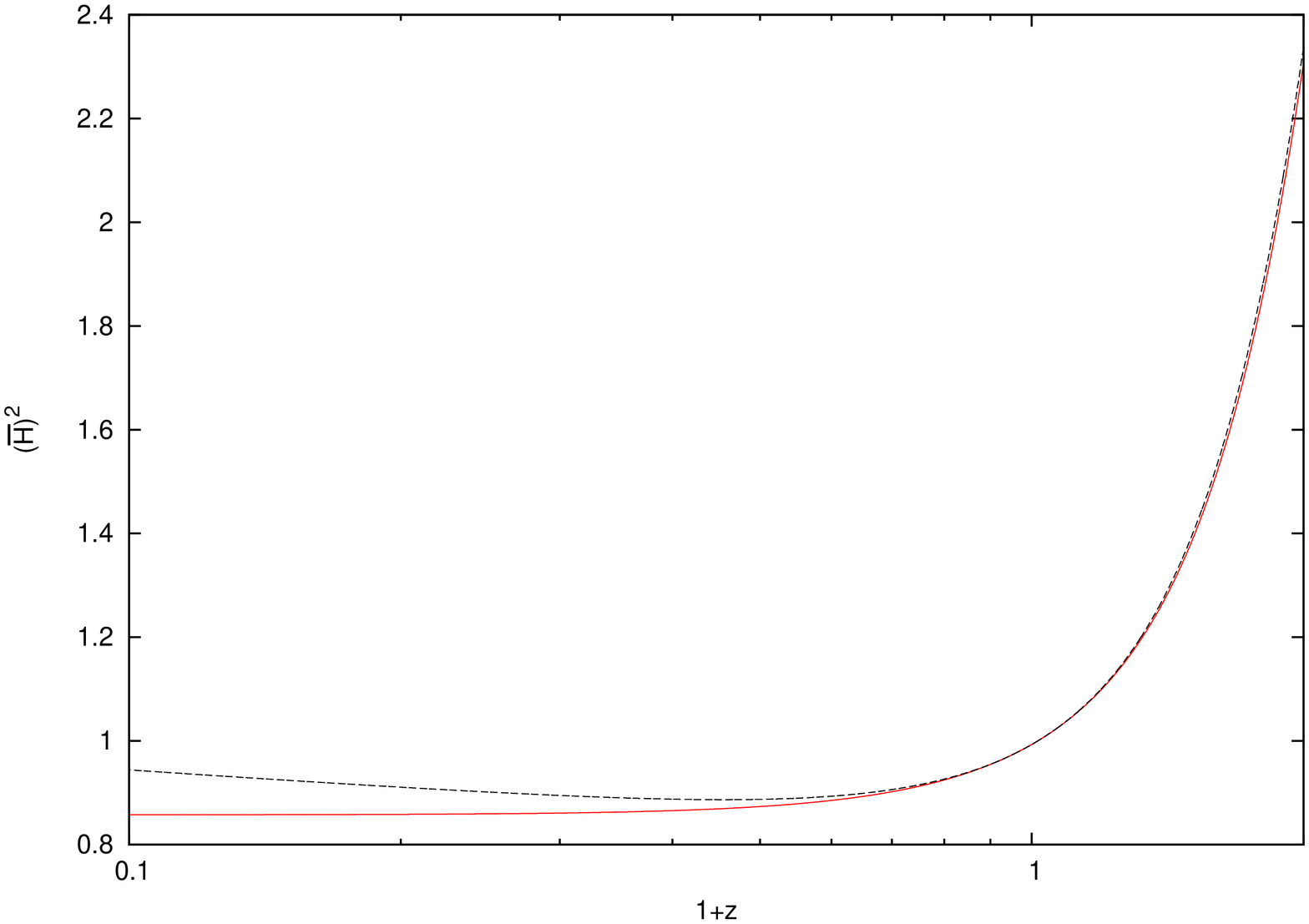}}} 
\mbox{\resizebox{0.48\textwidth}{!}{\includegraphics[angle=270]{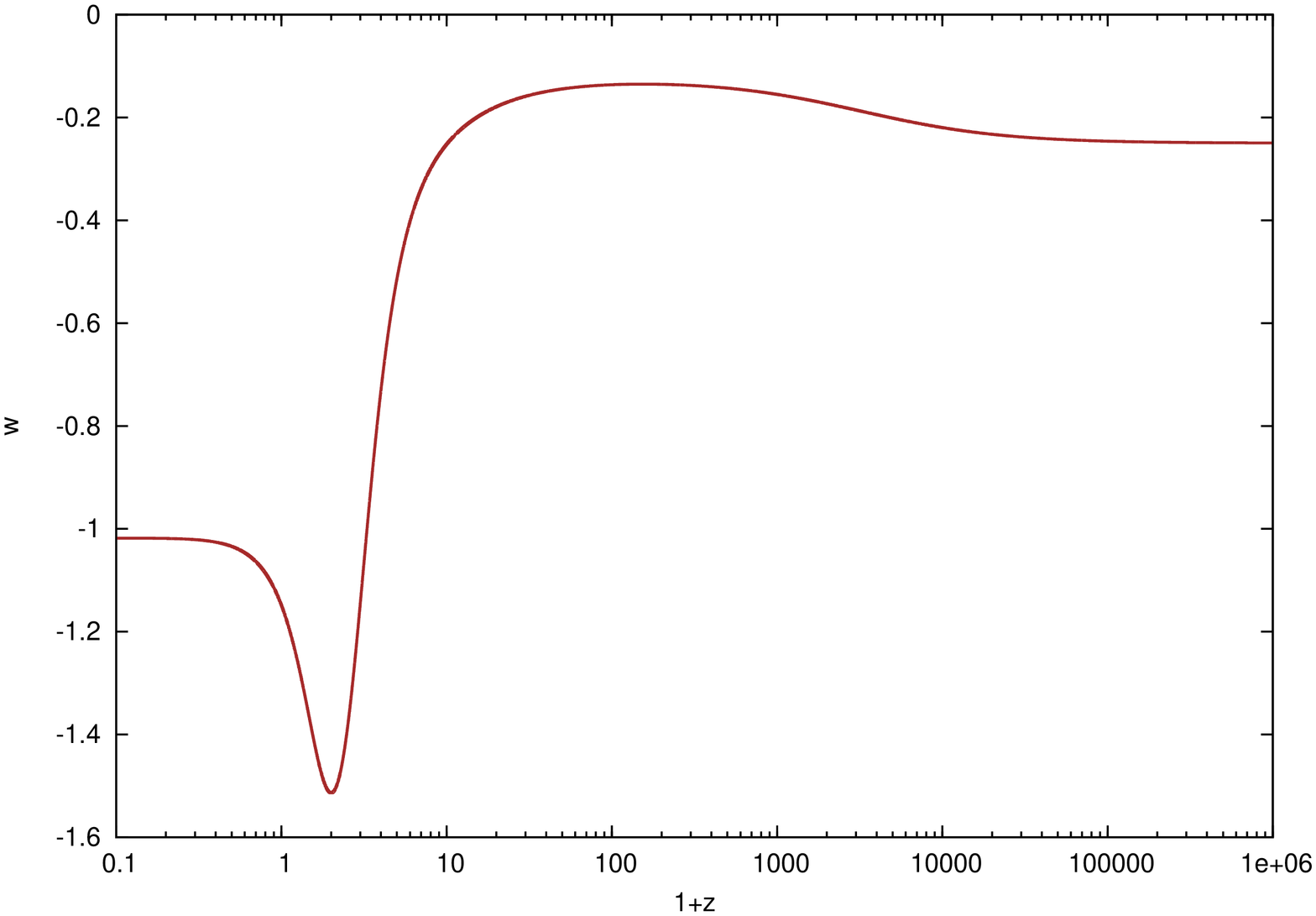}}} 
\caption{\label{fig:f19} [Left panel] 
The redshift evolution of viable linear coupled Galileons is restricted to 
be near that of the equivalent uncoupled Galileon, except at late times. 
The late time behaviour of $\bar{H}^{2}$ for the linearly coupled model 
(dotted black) is compared to the uncoupled model $\bar{H}^{2}$ (solid red) 
with the parameter 
choices $(c_{2},c_{3},c_{4},c_{5},c_{0}) =  (-5.9,-2,-1,-1,0.05)$ 
and $(-6.0,-2,-1,-1, 0)$ respectively ($c_{2}$ is adjusted to ensure 
$\Omega_{\pi,0} =0.76$ in both cases).  Note the Hubble parameter in the 
linear coupling case no longer asymptotes to a constant value.  
[Right panel] The effective equation of state for the linearly coupled 
$\pi$ field as a function of redshift.  At early times $w$ resembles the 
uncoupled case, e.g.\ Fig.~\ref{fig:expunc}, but deviates at later times 
and in the future has $w$ close to but not exactly $-1$. }
\end{figure}

The numerical evolution of the linearly coupled model is exhibited in 
Figure~\ref{fig:f19} for generic parameter values $c_{0} = 0.1$ $c_{3}=-1$, 
$c_{4} = 1$, $c_{5} = -1$, which fixes $c_{2} = -10.0$ 
so that $\Omega_{\pi}(z=0) = 0.76$. The high redshift evolution is 
necessarily similar to the uncoupled case, as just discussed, but the 
future evolution differs.  We note that $\bar{H}$ and $x$ 
continue to evolve in the future and do not approach constant values as 
in the uncoupled case, so there is no de Sitter state; nevertheless the 
equation of state $w$ remains very close to $-1$.

\subsection{Derivative Coupled Galileon: $c_G\ne0$, $c_0=0$} \label{sec:deriv}

In this subsection we include the derivative coupling term with $c_G$, 
and switch off the linear coupling term with $c_0$.  Recall that the 
kinetic coupling to the Einstein tensor, $G^{\mu\nu}\partial_\mu\pi 
\partial_\nu\pi$, also arises in some higher dimension gravity theories 
and disformal field theories and by itself has interesting behavior 
involving cosmic acceleration \cite{gubitosi,derham0610}. 

At early times, the $c_G$ contribution will dominate over the $c_2$ one, 
and for reasonable initial conditions (i.e.\ that the energy 
density $\rho_{\pi} \ll \rho_m$ initially) $x$ will satisfy $x \ll 1$ and 
hence the $c_G$ term will dominate over the $c_3$ term as well. 
However the importance of the $c_G$ 
term relative to $c_{4,5}$ depends on the specific initial conditions.  
If $c_G$ does not dominate, then we have the previous uncoupled case 
behaviors.  If $c_G$ does dominate then during radiation domination 
\begin{equation} 
\label{eq:i19}  c_{\rm G} : \qquad x \sim a^{3} \qquad \rho_{\pi} \sim 
c_{\rm G} a^{-2} \qquad w=-1/3  
\end{equation} 
and during matter domination 
\begin{equation} 
\label{eq:i18}  c_{\rm G} : \qquad x \sim a^{3/2} \qquad \rho_{\pi} \sim 
c_{\rm G} a^{-3}  \qquad w=0 \,. 
\end{equation} 

\begin{figure}
\centering
\mbox{\resizebox{0.48\textwidth}{!}{\includegraphics[angle=270]{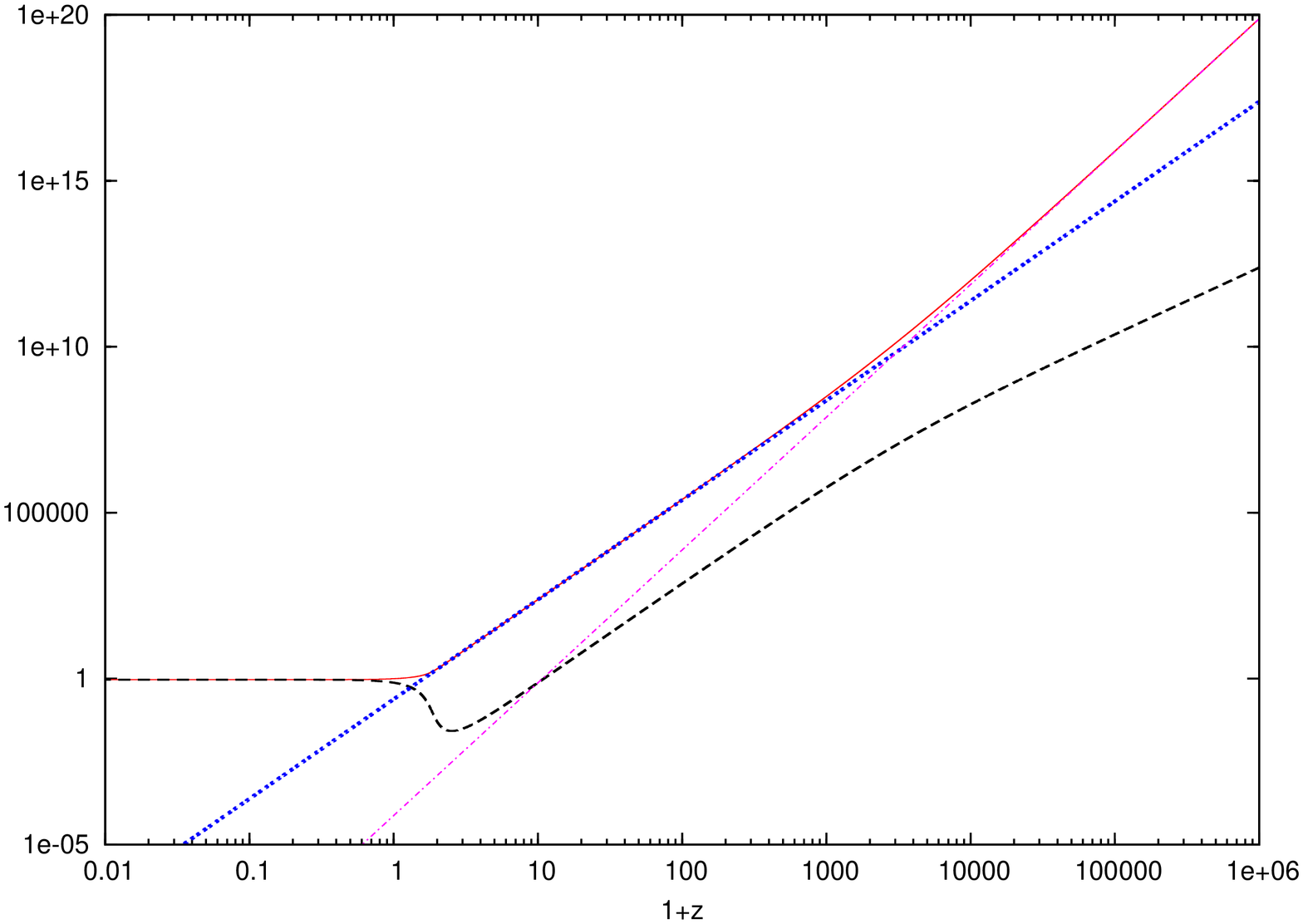}}}
\mbox{\resizebox{0.48\textwidth}{!}{\includegraphics[angle=270]{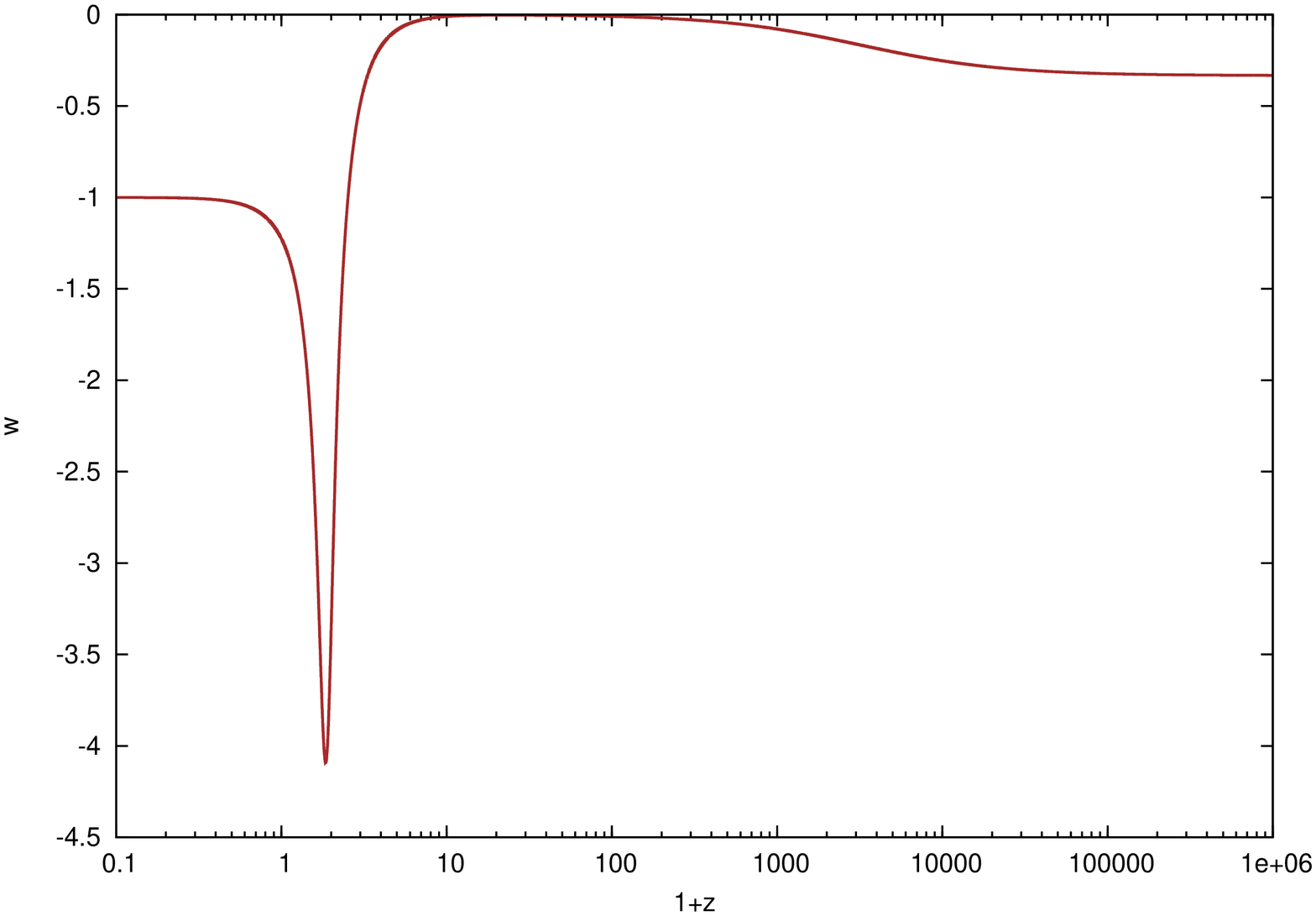}}}
\caption{\label{fig:f23} [Left panel] The evolution of $\bar{H}^{2}$ 
(solid red), $8\pi G \rho_m/3H_{0}^{2}$ (dotted blue), 
$8\pi G \rho_r/3H_{0}^{2}$ (dot-dashed magenta) 
and $8\pi G \rho_{\pi}/3H_{0}^{2}$ (long dash black) for the derivatively 
coupled Galileon, with $c_{2} = -9.1$, $c_{3} = -1$, $c_{4}=c_{5} =0$. 
[Right panel] The equation of state of the 
$\pi$ field as a function of redshift. The numerical evolution exhibits the 
analytic attractor results.} 
\end{figure}

Note that $\rho_{\pi}$ grows relative to the matter energy density during 
the radiation epoch and scales with it for $z \lesssim 10^{3}$, and 
therefore we must be careful to choose initial conditions such that 
$\rho_{\pi}$ remains subdominant until $z\sim 1$. In particular, we 
impose that $\Omega_{\pi} \lesssim 2\times 10^{-2}\, \Omega_m$ at 
matter/radiation equality.  We exhibit the full behaviour in 
Figure~\ref{fig:f23}, numerically evolving the equations for the case 
$c_{4} = c_{5} = 0$, $c_{3} = -1$ and $c_G = -1$.  The transition from 
$w=-1/3$ during radiation domination to $w=0$ for $z < 1000$ is clear.  
If we switch on the $c_{4,5}$ terms then since those terms grow 
relative to the $c_G$ term during radiation domination the limiting 
cases of Eqs.~ (\ref{eq:i19}) and (\ref{eq:i18}) will no longer be 
clearly distinguishable. 

At late times the derivative coupled Galileon goes to the de Sitter 
attractor.  Note that in the model of \cite{gubitosi} using derivative 
coupling, the $c_2$ term was forced to the canonical form, and so no 
stable de Sitter attractor was allowed, however here there is 
sufficient freedom to enable an asymptotic de Sitter state.

\section{Evolution of Perturbations and Gravitational Strength} \label{sec:geff} 

Having established the behaviour of the homogeneous cosmological expansion 
for the classes of Galileon models, we now examine the behaviour of 
linearized perturbations. Since the derivatives of the $\pi$ field couple 
to the metric potentials, we expect the field to modify the growth of 
matter density perturbations by introducing time dependent couplings between 
$\delta \pi$, $\phi$ and $\psi$. The equations governing the behaviour of 
subhorizon perturbations are given in Appendix~\ref{sec:applineq}. 
As discussed in Section~\ref{sec:pert} 
the effects on growth of structure (and light deflection) are concisely 
encapsulated in the gravitational strength modification functions 
$\geff^{(\psi)}$ and $\geff^{(\psi+\phi)}$. It has been shown in \cite{Babichev:2011iz,Kimura:2011dc} 
 that stringent constraints can be placed on the Galileon class of models 
 by considering the time evolution of $G_{\rm N}$.

The ``paths of gravity'' describe the evolution of these functions, from 
the initial deviation from the general relativistic value of $G_{\rm N}$ at 
high redshift to present day signatures and the eventual asymptotic values. 
It is important to stress the domain of validity of our equations; they are 
applicable to linear perturbations on subhorizon scales, assuming that dark 
matter and baryons can be treated as a common fluid.  We consider the 
evolution of perturbations over the redshift range $z < 500$.  We set 
initial conditions at $z = 10^{6}$ as $\rho_{\pi} = 10^{-5}\rho_m$, to 
ensure that at $z\sim 500$ the perturbations remain small and $G_{\rm eff}$ 
is sufficiently close to its general relativistic behaviour. At low redshifts, linear perturbation
theory is only applicable for large scale modes. To model non-linear perturbations
at the present time we would be forced to resort to simulations, which is beyond the 
scope of the present work. Significant deviations from the linearized behaviour will be expected
in the non-linear regime, due to the presence of a cosmological Vainshtein screening effect.

The perturbation equations for the $\pi$ field itself must also satisfy 
certain physical conditions for a sound theory, such as the absence of 
ghosts modes and Laplace instabilities.  These will constrain the allowed parameter 
space, as can conditions placed on the sign and magnitude of $\geff$.  
Before discussing the gravitational evolution we briefly discuss the 
no-ghost condition, later treating its details and the Laplace instability 
in Appendices~\ref{sec:appghost} and \ref{sec:applapl}.

\subsection{No-Ghost Condition} \label{sec:noghostmain} 

The equation of motion for the field perturbations takes the form of a 
wave equation, with the sign of the second time derivative $\delta\ddot y$ 
needing to be non-negative to avoid ghosts (negative kinetic terms causing 
the Hamiltonian to be unbounded from below).  For the standard, simple case 
of a minimally coupled, canonical scalar field the condition is just 
positivity of the kinetic energy, i.e.\ $c_2>0$ when all other $c$'s are 
zero.  In the presence of nonlinear kinetic terms in the action, and 
couplings, the situation is more complicated and we present our analysis 
in Appendix~\ref{sec:appghost}.  Here we summarize the result: to be free 
of ghosts the theory must satisfy 
\begin{equation}  
\label{eq:ghostmain} -{c_{2} \over 2} + 6c_{3} \bar{H}^{2}x + 3c_G\bar{H}^{2} 
- 27c_{4} \bar{H}^{4} x^{2} + 30 c_{5} \bar{H}^{6}x^{3} + 
2{\left( 3c_{3}\bar{H}^{2}x^{2} + 6c_G \bar{H}^{2}x 
- 18c_{4} \bar{H}^{4} x^{3}  + {45 \over 2} c_{5} \bar{H}^{6} x^{4}  
-3c_{0} \right)^{2} \over -6( 1 -2c_{0}y) - 6c_G \bar{H}^{2}x^{2} 
+ 9c_{4} \bar{H}^{4} x^{4} - 18c_{5} \bar{H}^{6} x^{5}} < 0 \,. 
\end{equation}

\subsection{\label{sec:early} The Thawing of Gravity at Early Times} 

At early times, with the Galileon contributions to the energy density 
small, the gravitational strength functions go to the general relativistic 
values of unity.  That is, the early universe behaves as in standard 
gravity and cosmology.  As the Galileon energy density increases, the 
modifications to gravity grow; we can say that the theory thaws away 
from general relativity, in analogy to the thawing class of dark energy 
dynamics (where the scalar field moves away from a frozen, cosmological 
constant state). 

For the initial stages of thawing we can calculate both $\geff$ and 
the no-ghost condition analytically for each of the cases where a given 
term of $c_{2-5,G}$ dominates the $\pi$ energy density. For the no-ghost 
condition, we find 
\begin{eqnarray} 
\label{eq:gh10} & & c_{5} : \qquad   - {\bar{\rho}_{\pi} \over \bar{H}^{2}} - {15 \over 56} \left( {\bar{\rho}_{\pi} \over \bar{H}^{2}}\right)^{2} < 0  \\ \label{eq:ghm1} & & c_{4} : \qquad -  {\bar{\rho}_{\pi} \over \bar{H}^{2}} - {8 \over 45} \left( {\bar{\rho}_{\pi} \over \bar{H}^{2}}\right)^{2} < 0  \\ \label{eq:gh2} & & c_{3} : \qquad  -  {\bar{\rho}_{\pi} \over \bar{H}^{2}} - {1 \over 12} \left( {\bar{\rho}_{\pi} \over \bar{H}^{2}}\right)^{2} < 0  \\ \label{eq:gh3} & & c_{2} : \qquad    - {\bar{\rho}_{\pi} \over \bar{H}^{2}}  < 0 \\   \label{eq:gh39} & & c_{\rm G} : \qquad  - {\bar{\rho}_{\pi} \over \bar{H}^{2}} - {4 \over 9} \left( {\bar{\rho}_{\pi} \over \bar{H}^{2}}\right)^{2} < 0  
\end{eqnarray} 
where $\bar{\rho}_{\pi} = \rho_{\pi}/(H_{0}^{2}M_{\rm pl}^{2})$ is the 
dimensionless Galileon energy density. In all cases, the condition 
${\rho}_{\pi} > 0$ is sufficient to ensure that the field is not a ghost.  
Note that given the initial conditions such that $\bar{\rho}_{\pi}  \ll 
\bar{H}^{2}$ during matter and radiation domination, the first terms in 
each will always dominate. 

\begin{figure}
\centering
\mbox{\resizebox{0.6\textwidth}{!}{\includegraphics[angle=270]{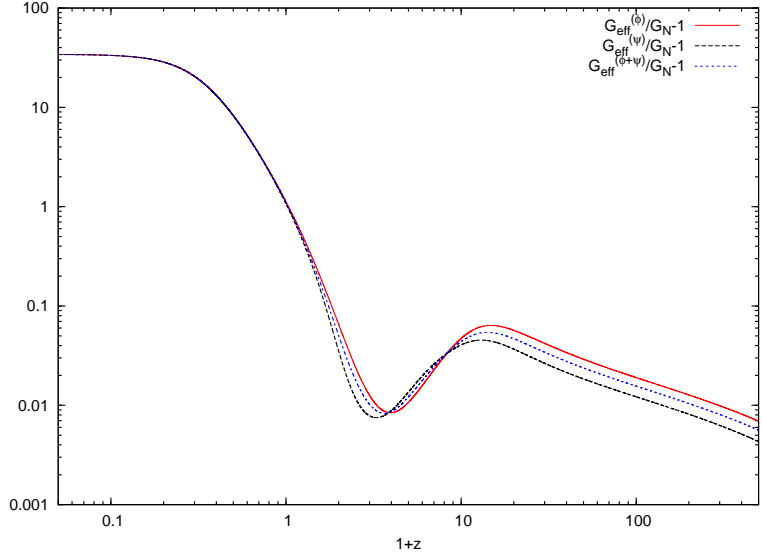}}}
\caption{\label{fig:gearly} The thawing of gravity away from its general 
relativistic strength, $G^{(\phi)}_{\rm eff}/G_{\rm N} - 1$ (solid red),  
$G^{(\psi)}_{\rm eff}/G_{\rm N} - 1$ (long dash black), and 
$G^{(\phi+\psi)}_{\rm eff}/G_{\rm N} - 1$ (short dash blue)
for model parameters $(c_{2},c_{3},c_{4},c_{5}) = (-15.3,-5.73,-1.2,-1)$.  
All three functions share a common redshift dependence at early times and
a common future asymptote, as discussed in the text. }
\end{figure}

For the gravitational strength, we find 
\begin{eqnarray} 
\label{eq:ge10} & & c_{5} : \qquad   {G^{(\phi)}_{\rm eff} \over G_{\rm N}} = 1 + {48 \over 7} \Omega_{\pi}     \\ \label{eq:ge1} & & c_{4} : \qquad  {G^{(\phi)}_{\rm eff}\over G_{\rm N}} = 1 +  \Omega_{\pi}     \\ 
\label{eq:ge2} & & c_{3} : \qquad   {G^{(\phi)}_{\rm eff} \over G_{\rm N}} = 1 + {1 \over 5} \Omega_{\pi}  \\ & & c_{2} : \qquad   {G^{(\phi)}_{\rm eff} \over G_{\rm N}} = 1    \\ & & c_{\rm G} : \qquad  {G^{(\phi)}_{\rm eff} \over G_{\rm N}} = 1 +  \Omega_{\pi}  \ . 
\end{eqnarray} 
These analytic solutions are only valid during matter domination. 
Figure~\ref{fig:gearly} exhibits the evolution of the deviations from 
general relativity $\geff/G_N-1$ for each of the effective gravitational 
strengths, from an early time where they share similar 
redshift evolution in the deviation from general relativity, to a shared 
late time de Sitter asymptote as discussed in the following section.

\subsection{\label{sec:late} Gravity in the de Sitter Limit}

The other limit of importance is the late time approach to the de Sitter 
attractor that is present in the uncoupled and derivatively coupled models.  
In the de Sitter limit, we have $\bar{H}', x' \to 0$, and it is convenient 
to define 
\begin{eqnarray} \label{eq:C1}
& & C \equiv c_{2} x^{2} \\ \label{eq:D1} & & D \equiv c_{3} \bar{H}^{2} x^{3} \\ & & E \equiv c_{4} \bar{H}^{4} x^{4} \\ & & F \equiv c_{5} \bar{H}^{6} x^{5} \\ & & A \equiv c_{G} \bar{H}^{2} x^{2} 
\ . 
\end{eqnarray} 
Note that generically all these terms are of the same order (unless a 
coefficient is zero) so we cannot study the limit in terms of a single 
dominant component, making the analysis more difficult but the behavior 
richer. 

In general (with one type of coupling) we have five parameters and two 
independent equations of motion.  The background field equations then 
reduce to algebraic constraints: 
\begin{eqnarray} 
\label{eq:l1} & & E = {1 \over 9} \left( -2A+8D-3C-10 \right) \\ 
\label{eq:lo2} & & F = {2 \over 3} \left( D - A - {C \over 2} - 2 \right) \ . 
\end{eqnarray} 
Eliminating $E,F$ from $G^{(\phi)}_{\rm eff}$, we arrive at the de Sitter 
limit 
\begin{equation} 
\geffds = -{3 \over 14+7A+3C-4D}\ G_{\rm N} \ . 
\end{equation} 
Interestingly, in the de Sitter limit $\geff^{(\phi)}=\geff^{(\psi)}= 
\geff^{(\psi+\phi)}$ -- a special property of the Galileon models that 
in this limit they have no gravitational slip -- 
and we write them simply as $\geffds$. 

It is also useful to write the no-ghost condition at the de Sitter 
state (although the no-ghost condition must hold at all times) in terms 
of $A,C,D$.  We find 
\begin{equation} 
\label{eq:ng1} -{3 \over 2} C + 2D - 11A -10 + 
2{\left( - {3 \over 2} C + 2D - 5A -10\right)^{2} \over 8+3C-4D+4A } < 0 \ . 
\end{equation} 
For the uncoupled case $A=0$, the no-ghost condition corresponds to 
\beq 
2+\frac{3C}{4}<D<5+\frac{3C}{4} \ . 
\eeq 

\begin{figure}[htbp!]
\includegraphics[width=0.48\textwidth]{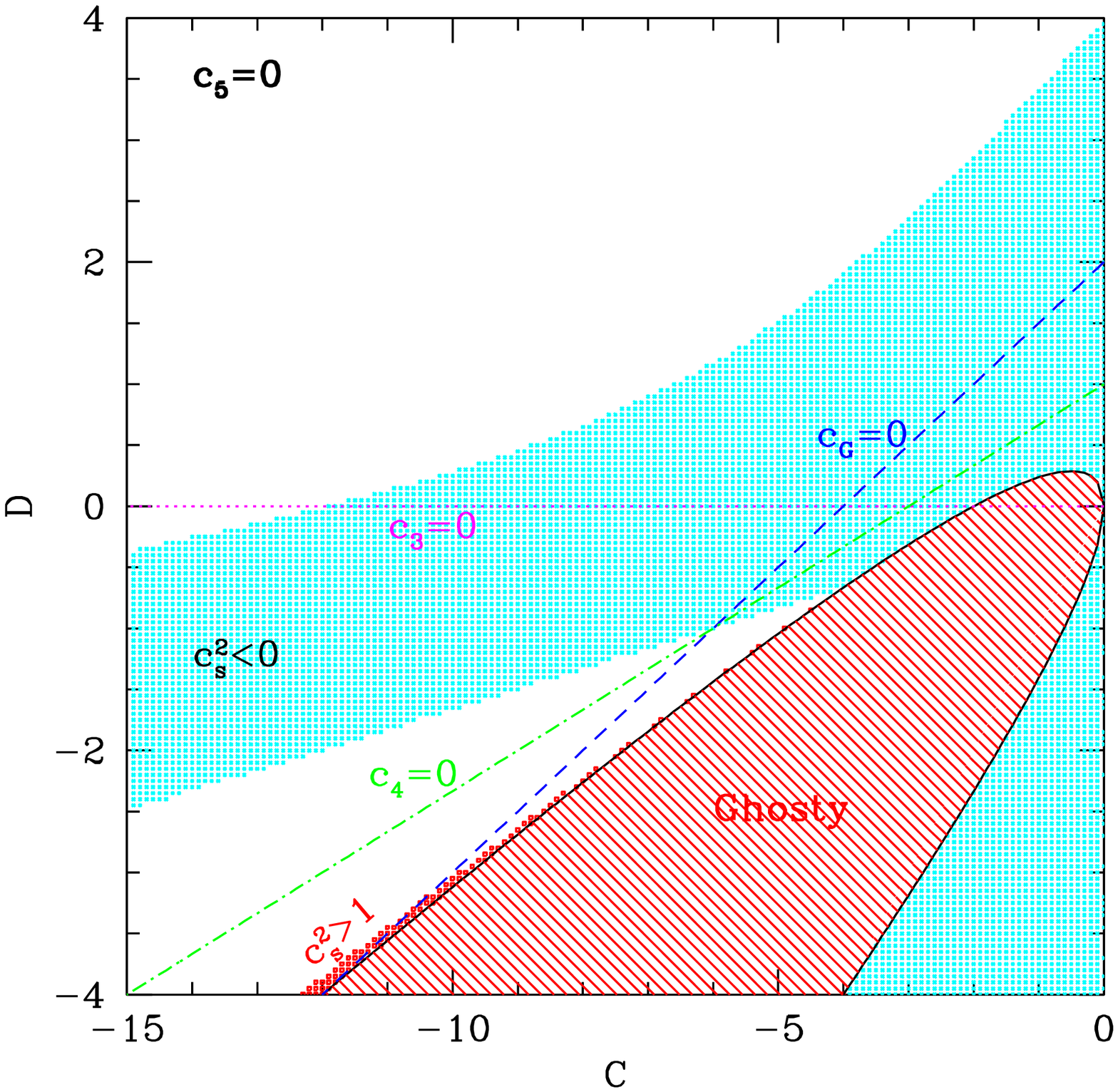}
\includegraphics[width=0.48\textwidth]{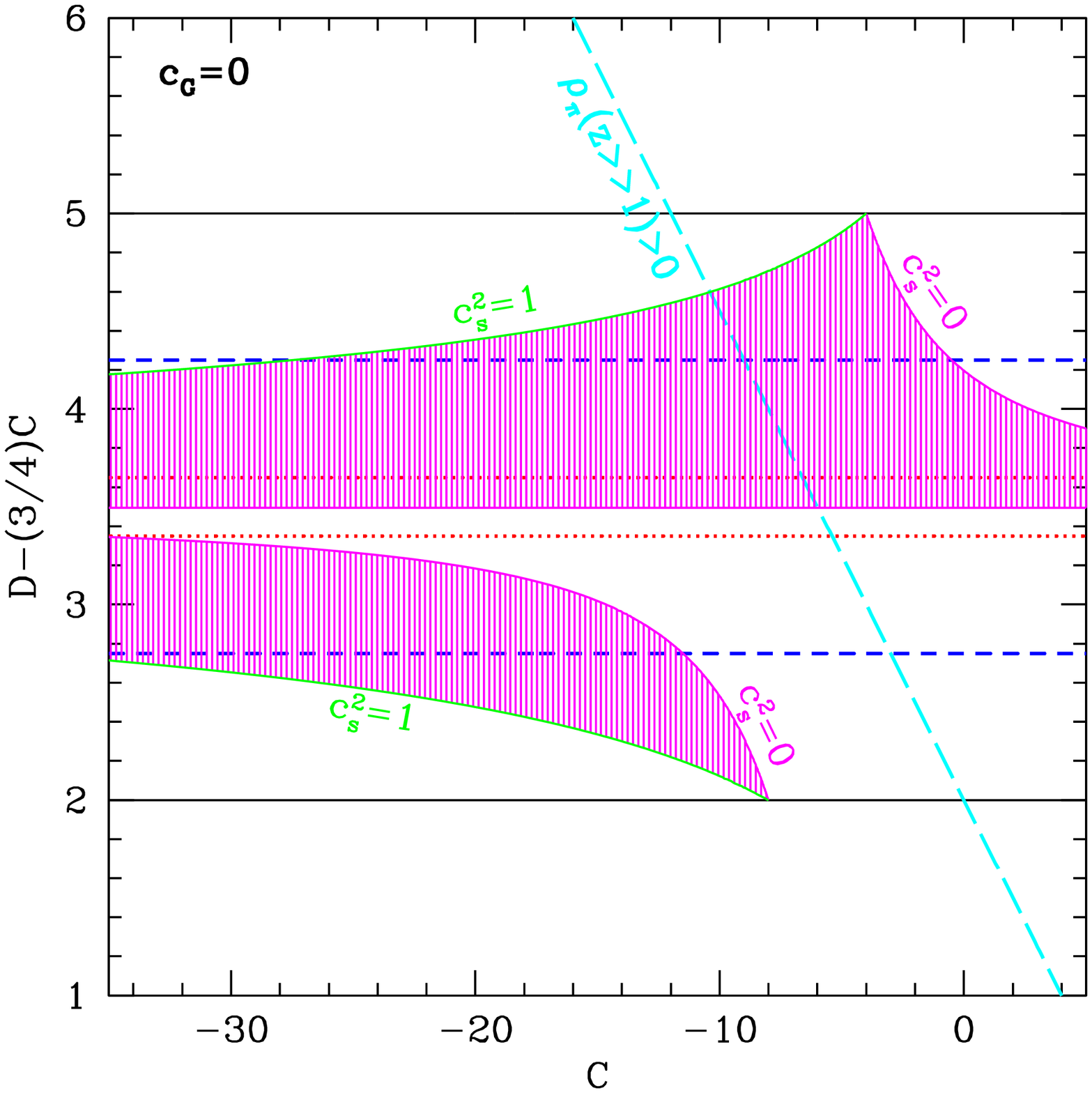}
\caption{[Left panel] Dark red shaded region shows the area violating the 
no-ghost condition in the $C$-$D$ plane, fixing $c_5=0$.  The straight 
lines show the cuts through the plane corresponding to special values 
$c_n=0$ as labeled.  The light cyan shaded 
region shows the Laplace instability $c_s^2<0$, and the thin ``bubbly'' red 
region shows the superluminal condition $c_s^2>1$ (see 
Appendix~\ref{sec:applapl}).  $C>0$ is also ruled out. 
[Right panel] For the $c_G=0$ case the no-ghost region lies between the solid 
black lines.  These boundaries correspond to the limits $\geffds/G_N=\pm1/2$; 
the blue dashed lines give $\geffds/G_N=\pm1$, and the red dotted lines give 
$\geffds/G_N=\pm5$.  Positive values are for the upper line of each set. 
$\geffds$ passes through infinity for $4D-3C=14$.  Note $\geffds/G_N$ cannot 
lie between $-1/2$ and $1/2$.  Further constraints include positivity of 
scalar field energy density at high redshift, limiting the parameter space 
to the left of the long dashed, diagonal cyan line when the $c_5$ term 
dominates, and the Laplace inequality (\ref{eq:unclaplace}), with shaded 
magenta regions obeying $0\le c_s^2\le 1$. 
}
\label{fig:ghostcd}
\end{figure}

Figure~\ref{fig:ghostcd} illustrates the constraints imposed on the Galileon 
parameter space by several conditions for physical viability, including 
the no-ghost condition, Laplace stability condition on the sound speed, 
and positivity of 
energy density $\rho_\pi$, and their relation to the asymptotic de Sitter 
value of the gravitational strength $\geffds$. 
For the uncoupled case note that $c_{5} < 0$, which is generally required 
in the early universe barring fine tuning, requires that both $c_{3}$ and 
$c_{4}$ must also be negative.

Since after the equations of motion are applied one has $\geff$ as a 
function of three free parameters, it is difficult to make general 
analytic statements.  We begin by considering restricted cases, and then 
carry out a more general scan through the full parameter space.

\begin{figure}[htbp!]
\mbox{\resizebox{0.406\textwidth}{!}{\includegraphics{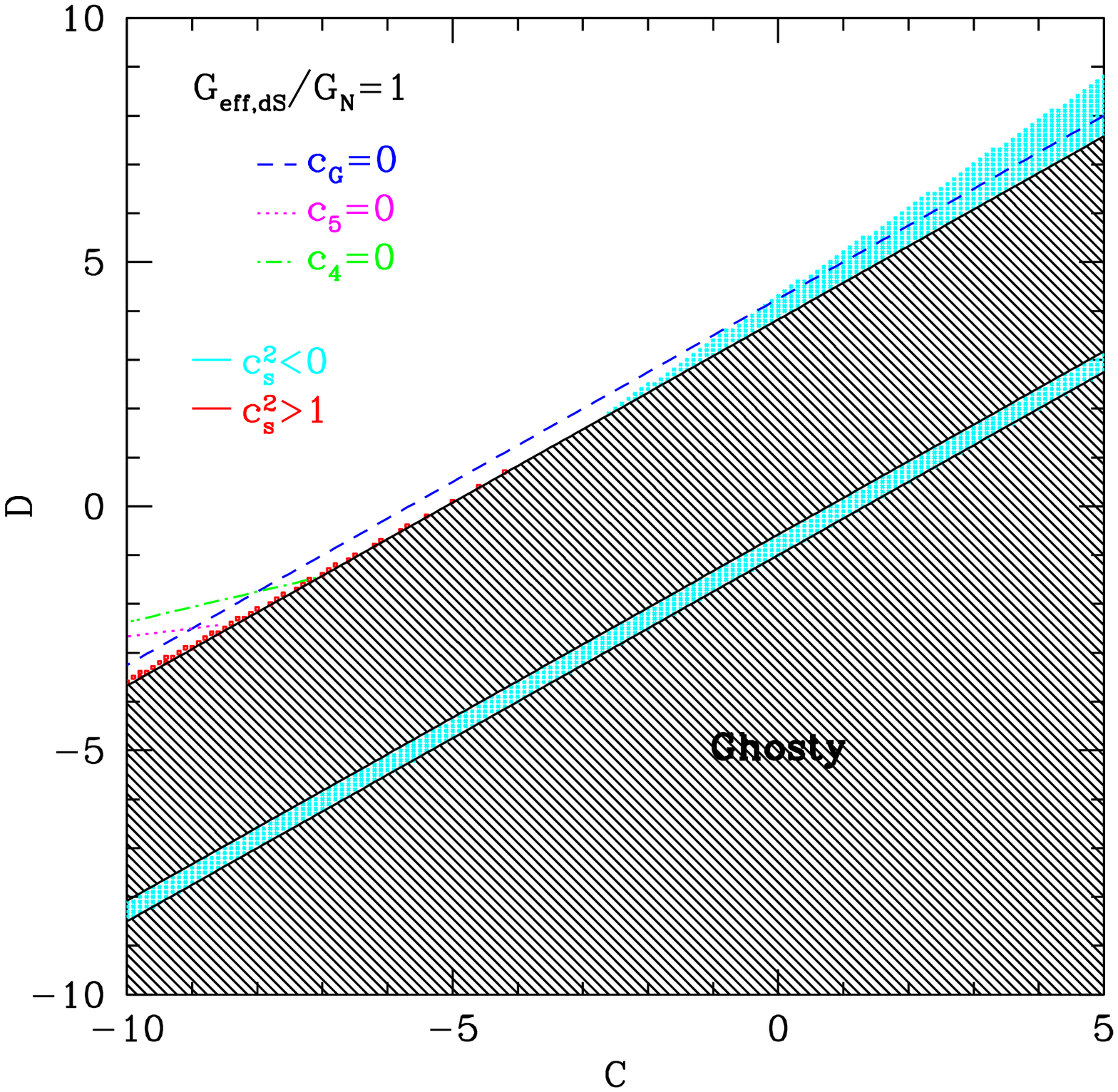}}}
\mbox{\resizebox{0.534\textwidth}{!}{\includegraphics[angle=0]{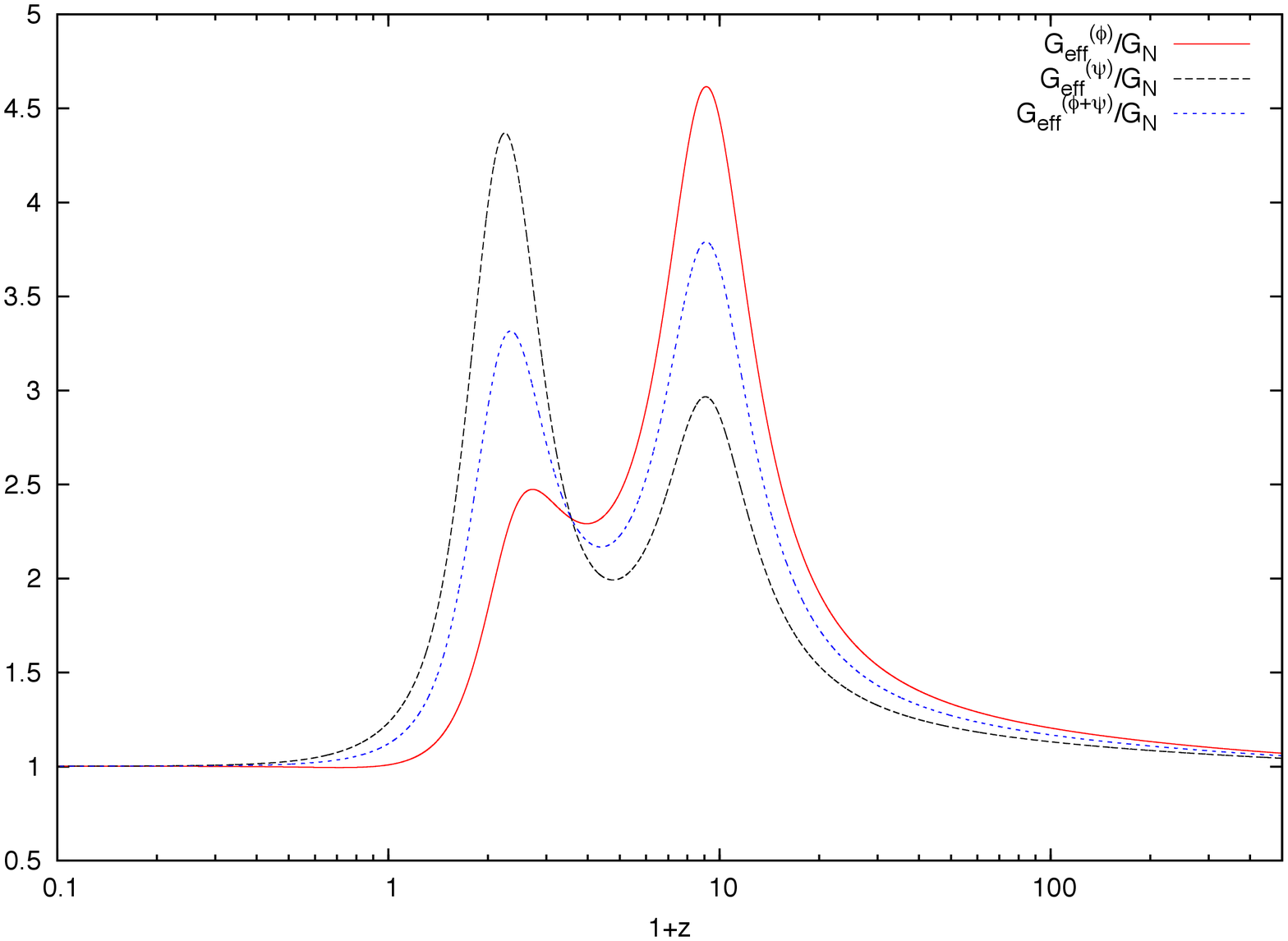}}} 
\caption{[Left panel] 
Region of the Galileon parameter space that gives a de Sitter 
limit of $\geff/G_N=1$ and has no ghosts or Laplace instability is in 
the unshaded region.  
The 2 dimensional viable region becomes a 1 dimensional line if further 
restrictions are imposed.  Note the uncoupled case $c_G=0$ is here always 
ghost free, though it is 
Laplace unstable, $c_s^2<0$, for insufficiently negative $C$. 
[Right panel] 
The evolution of the effective gravitational strengths 
$G_{\rm eff}^{(\phi)}$ (solid red), $G_{\rm eff}^{(\psi)}$ (long dash black) 
and $G_{\rm eff}^{(\phi+\psi)}$ (short dash blue) for an uncoupled Galileon 
case restoring to general relativity in the de Sitter limit, with 
and $(c_{2},c_{3},c_{4},c_5) = (-12.688,-6.239,-2.159,-1)$.  
Note the strengths can be significantly different from each other.
}
\label{fig:geffads1}
\end{figure}

First suppose that only one of the parameters $c_{2-5,G}$ is nonzero.  
We find that in no case is there a consistent de Sitter solution.  
Solutions are possible when, e.g., two or three $c_{2-5,G}$ terms are 
nonzero.  For a more physically motivated restriction, 
one can look at parameter choices for which gravity returns 
asymptotically to general relativity, at least in the sense of $\geffds=1$.  
This requires the condition 
\beq 
17+7A+3C-4D=0 \,, 
\eeq 
so now our 5 dimensional parameter space is restricted to 2 dimensions 
and we can plot allowed regions, imposing as well the viability conditions. 
The no-ghost condition becomes 
\beq 
3C-4D<\frac{-39-7\sqrt{57}}{6} \qquad {\rm or}\qquad 4>3C-4D> 
\frac{-39+7\sqrt{57}}{6} \,, 
\eeq 
(although the second region violates Laplace stability, having $c_s^2<0$) 
and is exhibited in the left panel of Figure~\ref{fig:geffads1}.  Setting 
some $c_n=0$ collapses the 2 dimensional space to a 1 dimensional line.

\subsection{The Paths of Gravity} \label{sec:paths} 

Having constructed some limiting cases analytically, we now examine 
the full parameter space by scanning through it, solving for $\geff$ 
numerically. Our aim is to find generic trends in the behaviour of 
$G^{(\phi)}_{\rm eff}$ and $G^{(\psi)}_{\rm eff}$, and furthermore 
restrict the parameter space by applying theoretical constraints on 
the behaviour of the $\pi$ field.

As an initial illustration of the paths of gravity, i.e.\ the evolution 
of the gravitational strength $\geff(a)$, we consider the family of models 
that result in $\geffds/G_N=1$.  For the uncoupled case $c_G= c_{0} =0$ our 
approach is to fix $c_{5}=-1$, and $c_{2}$ by demanding $\Omega_{\pi,0}=0.76$. 
There remains a set of $(c_{3},c_{4})$ values that can satisfy our remaining 
condition $\geffds/G_N=1$.

Figure~\ref{fig:geffads1}, right panel, plots the effective gravitational 
strengths 
$G_{\rm eff}^{(\phi)}(a)/G_{\rm N}$, $G_{\rm eff}^{(\psi)}(a)/G_{\rm N}$, 
and $G_{\rm eff}^{(\phi+\psi)}(a)/G_{\rm N}$ for one case giving rise to 
the asymptotic behaviour $\geffds/G_N=1$.  Despite having past and future 
asymptotes agreeing with general relativity, the intermediate evolution 
over the redshift range $z\approx(0,10)$ can be strongly distinct, both 
from general relativity and among the different $\geff$'s.  
The magnitude of the features over this range is extremely sensitive to 
the values of $c_{3,4}$.  

While it is instructive to analyze particular examples of model parameters, 
we can also learn a lot by a broader if shallower study. 
Generically we have a three (uncoupled case) or four (linearly or 
derivatively coupled cases) dimensional parameter space spanned by 
$c_{3-5}$, plus possibly $c_0$ or $c_G$.  As before we fix $c_2$ by 
requiring the present dark energy density $\Omega_{\pi,0} = 0.76$.  
In what follows we also fix $c_{5} = -1$ for tractability, and hence as 
mentioned in the previous subsection only have to consider the range 
$c_{3},c_{4} < 0$.  
We systematically scan the parameter space by selecting 22,500 points from 
a uniform grid over the range $(-5,0)$ for each of the $c_{3,4}$ terms.  
We discuss the $c_0$ and $c_G$ parameters below.  For each point we evolve 
the background equations ($\ref{eq:n8},\ref{eq:n9}$) from $z = 10^{6}$ to 
$z=-0.99$ (i.e.\ future scale factor $a=100$).  
From the 22,500 expansion histories, we keep as viable only those 
that satisfy the following conditions for $z < 500$: 

\begin{enumerate}[(a)] 

\item The energy density of the $\pi$ field is positive definite, 
$\rho_{\pi} > 0$. 

\item The no-ghost condition ($\ref{eq:ghost}$) is satisfied. 

\item The Laplace inequality ($\ref{eq:p1}$) is satisfied: 
$ c_s^2 \ge 0$.  

\end{enumerate}

The positivity of the energy density is not strictly required at all times during the cosmological history.
However, there are a number of reasons as to why we include it. We have already seen in ($\ref{eq:gh10}-\ref{eq:gh39}$)
that at early times the no-ghost constraint is intimately tied to the $\rho_{\pi}>0$ condition. Similarly, we must have $\rho_{\pi} > 0$ at 
the present time since we demand that the field drives the current late time acceleration. This does not preclude the possibility that
the energy density can be positive at high redshift, cross through $\rho_{\pi}=0$ at some intermediate time and then return to a positive
value at $z=0$. It is very difficult to prove that such a crossing can never occur, however we have reason to believe that such behavior
is highly unlikely. There are multiple fixed points that the galileon field can approach asymptotically; in addition to the 
de Sitter point there are both anti de Sitter and Minkowski vacuum states, and it is likely that any crossing of the zero line $\rho_{\pi}=0$ will result in the galileon failing to approach the de Sitter state that we require. Hence we impose $\rho_{\pi} > 0$.

\begin{figure}
\centering
\mbox{\resizebox{0.48\textwidth}{!}{\includegraphics[angle=270]{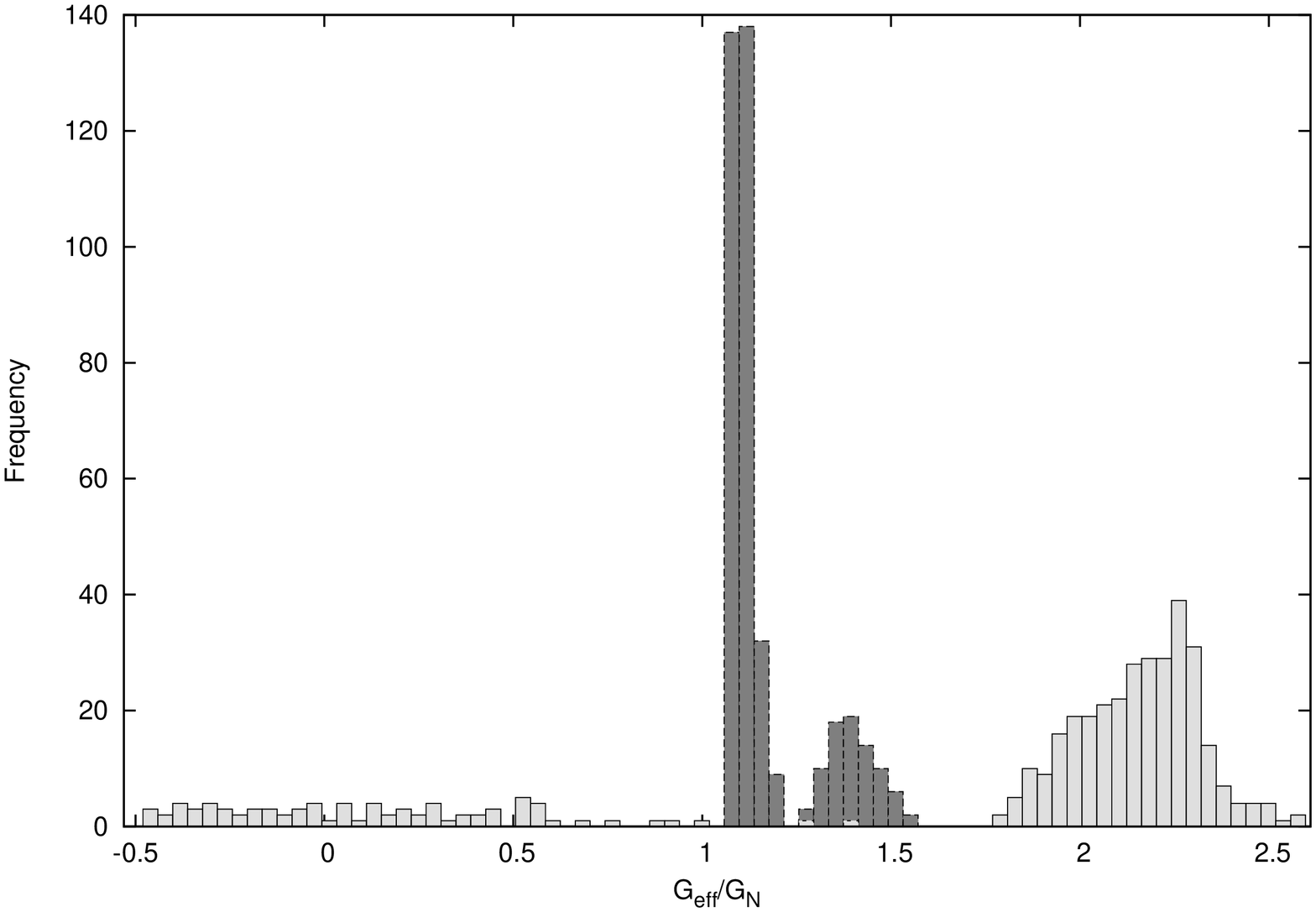}}}
\mbox{\resizebox{0.48\textwidth}{!}{\includegraphics[angle=270]{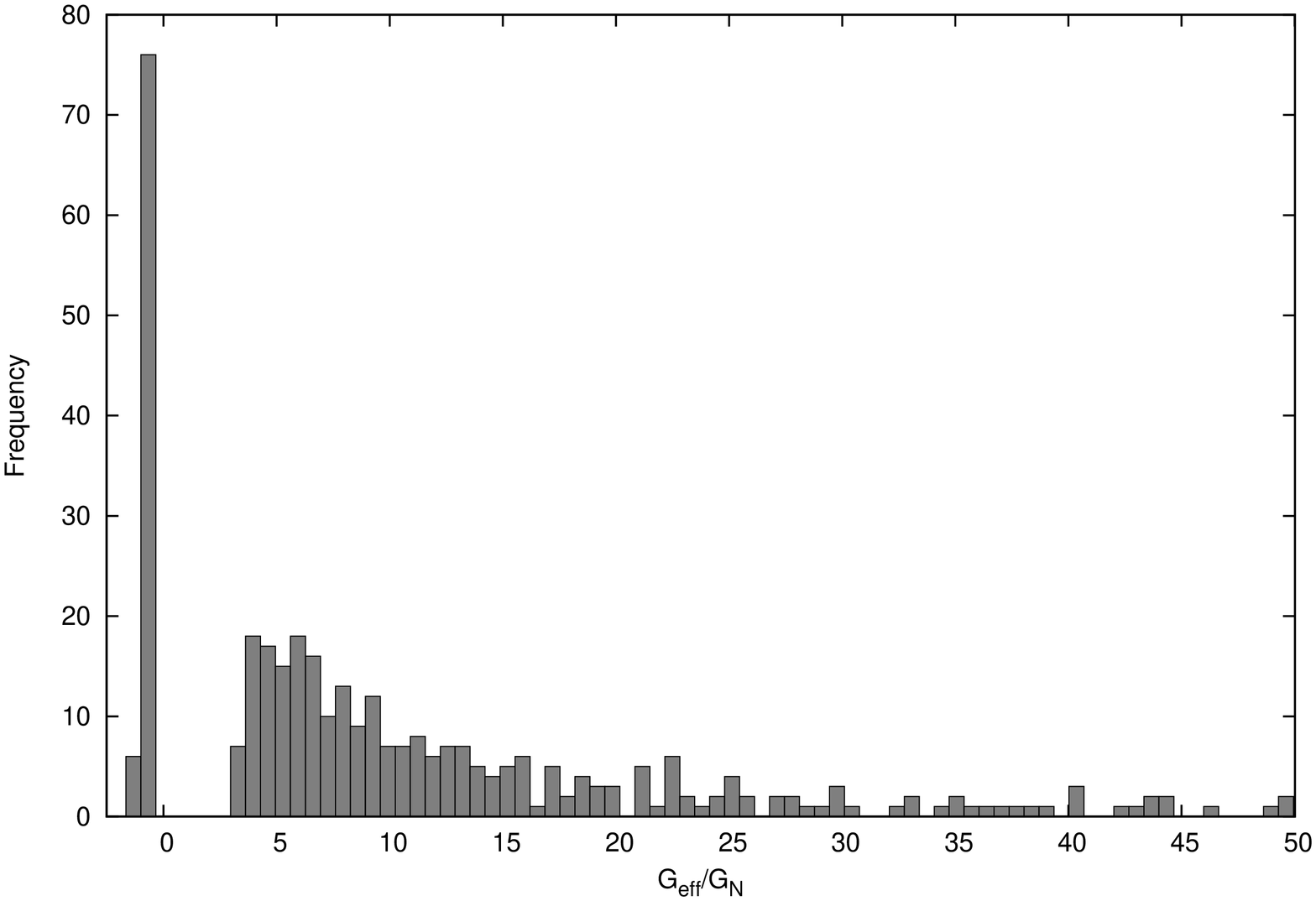}}}
\caption{\label{fig:hun} [Left panel] Histogram of the gravitational strength 
$G^{(\phi)}_{\rm eff}/G_{\rm N}$ 
for an ensemble of theoretically viable models arising from a uniform scan 
through the uncoupled Galileon parameter space, at redshift $z=1$ (dark) and 
$z=0$ (light).  At $z=1$ we find $G_{\rm eff}$ to be clustered in the 
vicinity of its general relativistic limit (deviating by $\sim10\%$), 
however at late times the behaviour is more diverse and further from the 
GR limit. We observe that $G_{\rm eff}$ can be both positive and negative at late times, although
the majority of the parameter space surveyed approached $G_{\rm eff} > 0$. [Right panel] The 
asymptotic value of $\geff$ at $z=-0.99$ (essentially the de Sitter asymptote), 
showing extremely diffuse behaviour in the future. } 
\end{figure}

Figure~\ref{fig:hun} plots the histograms of the results for the 
gravitational strengths $G^{(\phi)}_{\rm eff}$ at redshifts $z=0$, $z=1$, 
and $z=-0.99$ for each of the surviving, theoretically viable runs.  At 
$z=1$ the viable models are strongly clustered in the vicinity of 
$G^{(\phi)}_{\rm eff}/G_N \gtrsim 1$, with deviations from the high redshift, 
general relativistic (GR) value by only $\sim10\%$.  However by $z=0$, 
when cosmic acceleration is strong, gravity generically is quite distinct 
from GR and a diversity of behaviors is possible.  At late times, essentially 
at the de Sitter state, the value of the gravitational strength is diverse, 
and takes both positive and negative values of $G_{\rm eff}^{(\phi)}/G_N$ 
(cf.\  the right panel of Figure~\ref{fig:ghostcd}). 
If we were to fix $\geffds/G_N=1$, say, then we find much greater diversity 
at $z\approx1$ but of course uniformity in the future.

\begin{figure}
\centering
\mbox{\resizebox{0.48\textwidth}{!}{\includegraphics[angle=270]{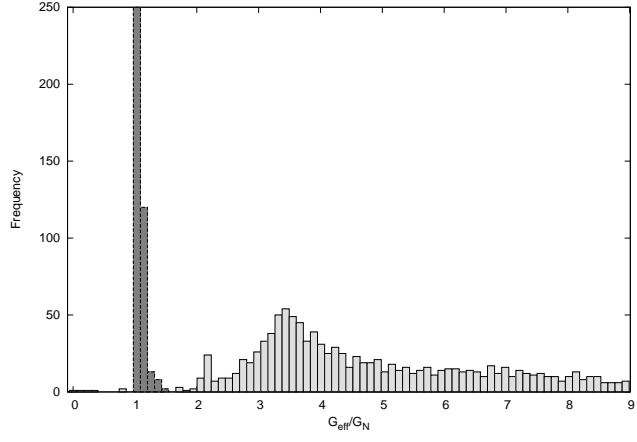}}}
\caption{\label{fig:Geffl} Histogram of the gravitational strength
$G^{(\phi)}_{\rm eff}/G_{\rm N}$ for an ensemble of theoretically viable 
models arising from a uniform scan through the linearly coupled Galileon 
parameter space, at redshift $z=1$ (dark) and $z=0$ (light).  We do not here 
consider models that diverge in the future to be unviable, so greater 
diversity is exhibited compared to the uncoupled model. } 
\end{figure}

We next perform an analogous numerical study for the linearly coupled 
Galileon model, expanding our parameter grid. We continue to fix $c_{5}=-1$, 
but now consider $250,000$ expansion histories over the range $c_{3}=(-5,5)$, 
$c_{4} = (-5,5)$ and $c_{0} = (-5,5)$.  The results are exhibited in 
Figure~\ref{fig:Geffl}. In this case, we no longer demand that our 
theoretical constraints are satisfied at all times, but rather only for 
$z > 0 $. We do not consider $z<0$ as there is no de Sitter fixed point for 
this model, so we would have to check infinitely far to the future, 
and $z=0$ is a logical cutoff in choosing a domain of applicability. 

The distribution of $G_{\rm eff}^{(\phi)}$ is similar to the uncoupled 
case -- recall from Sec.~\ref{sec:lin} that generally the $c_0$ contribution 
has to be subdominant at early times.  However the later time behaviour is 
more diffuse. This is due to the 
fact that we now permit $G_{\rm eff}$ to diverge for $z < 0$.  Such 
divergences can occur for any $z < 0$, and the value of $G_{\rm eff}$ at 
$z=0$ is particularly sensitive to the existence and redshift of these 
features.

\begin{figure}
\centering
\mbox{\resizebox{0.48\textwidth}{!}{\includegraphics[angle=270]{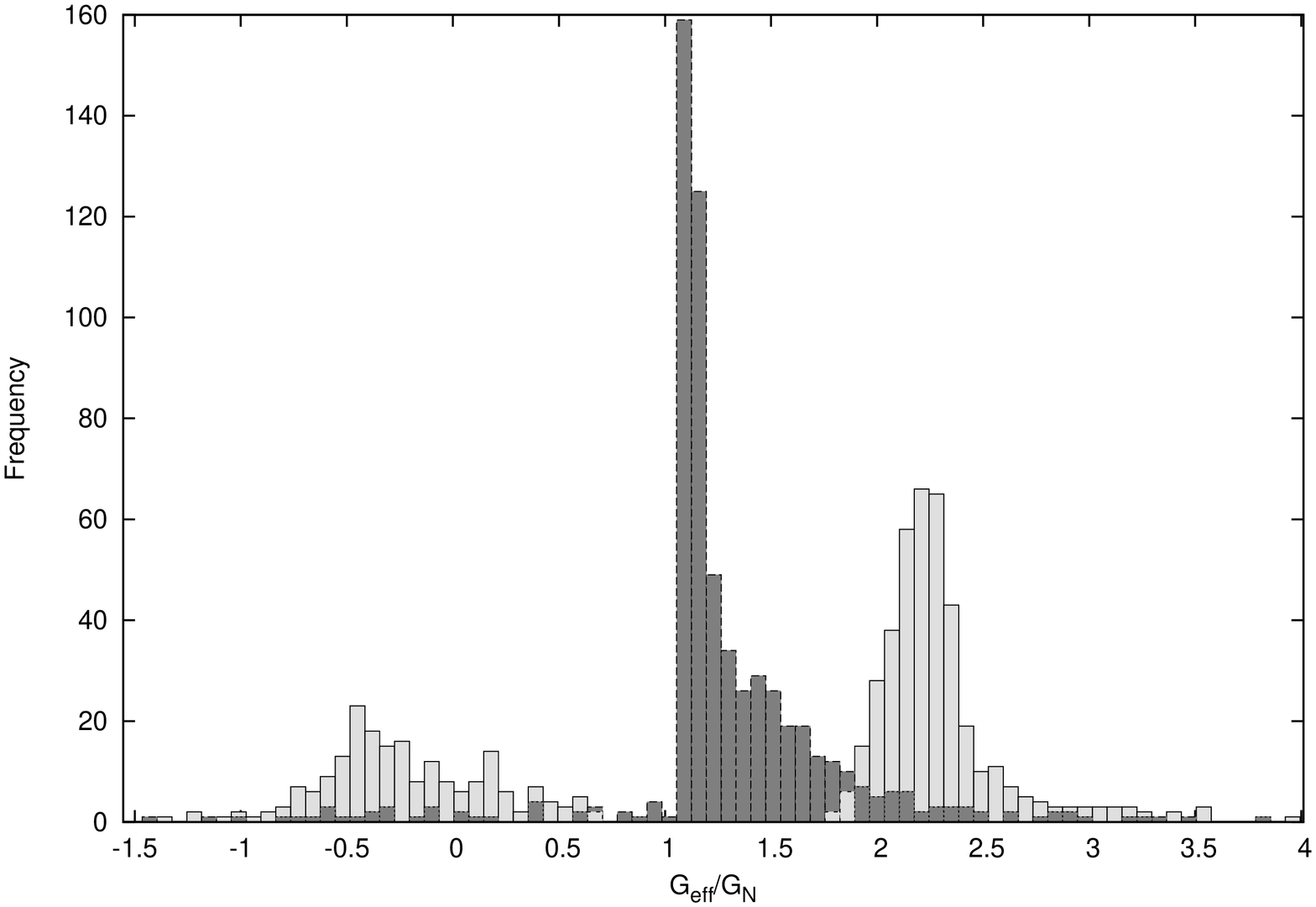}}}
\mbox{\resizebox{0.48\textwidth}{!}{\includegraphics[angle=270]{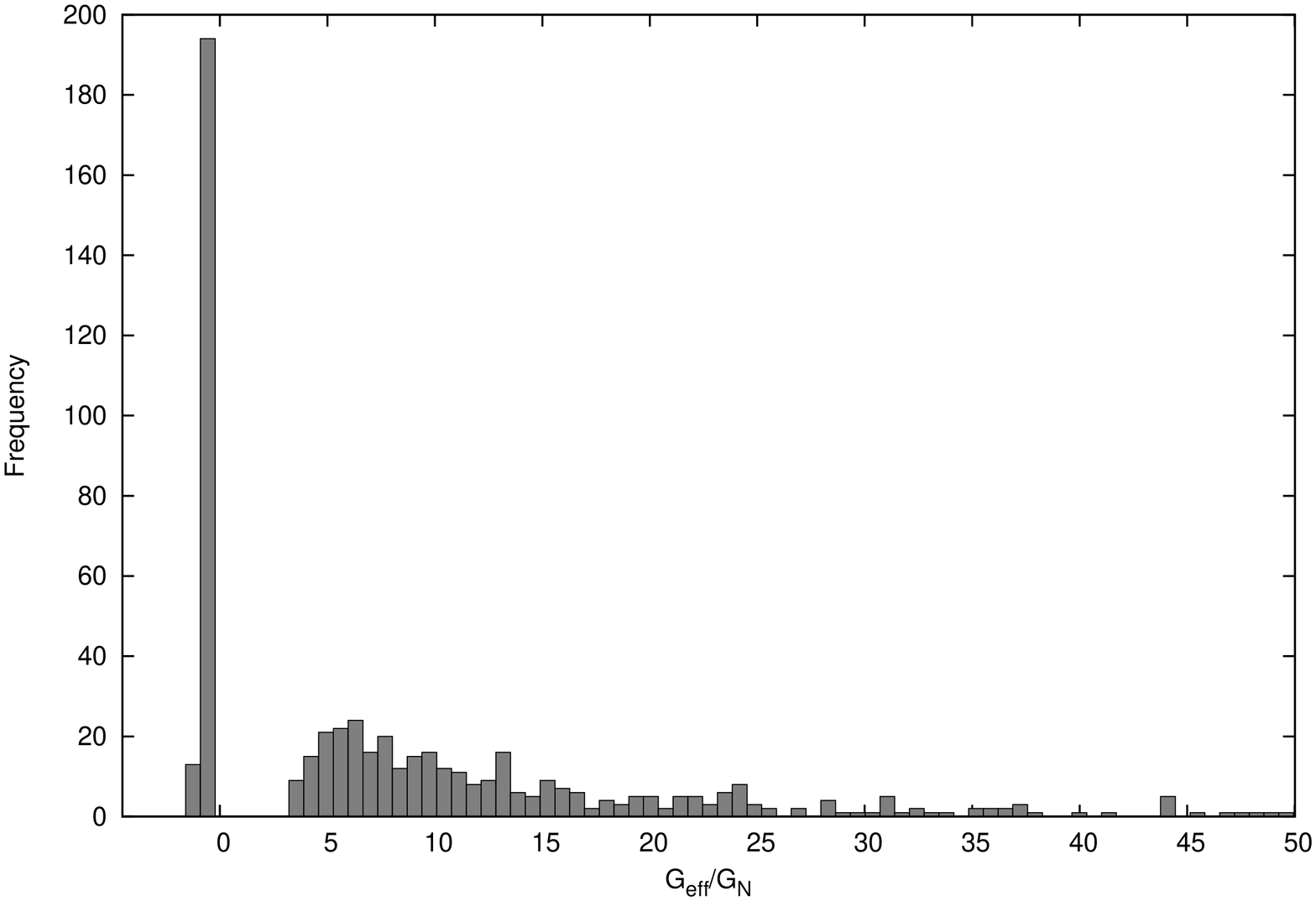}}} 
\caption{\label{fig:histd} Histogram of the gravitational strength
$G^{(\phi)}_{\rm eff}/G_{\rm N}$ for an ensemble of theoretically viable 
models arising from a uniform scan through the derivatively coupled Galileon 
parameter space, at redshift $z=1$ (dark) and $z=0$ (light) in the left panel, 
and $z=-0.99$ in the right panel.  We observe similar, although more 
diffuse, behaviour in the model with $c_{\rm G} \neq 0$ 
compared to the uncoupled case.} 
\end{figure}

Finally, we repeat our analysis for the derivatively coupled Galileon model, 
expanding our parameter grid to include $c_{\rm G} \ne 0$.  We evenly 
distribute 250,000 points amongst the three dimensional space 
$c_{3} = (-5,5)$, $c_{4} = (-5,5)$, $c_{\rm G} = (-5,5)$. The resulting 
histograms of theoretically viable models are 
exhibited in Figure~\ref{fig:histd}.  We observe qualitatively similar behaviour to the uncoupled model at both $z = 1$ and $z=0$, although we find considerably more diffuse behaviour at both $z=0$ and $z=1$, with a higher propensity towards negative asymptotes.

Over the course of our numerical study, we considered the range
$c_{\rm G} = (-5, 5)$. However we found no models which were deemed viable 
for $c_{\rm G} < 0$.  
The theoretical constraint that causes the derivatively coupled model to 
fail in this regime is the Laplace condition of Appendix~\ref{sec:applapl}, 
$c_{\rm s}^{2} > 0$.  If we evolve our equations back 
to radiation domination in 
the early universe, whenever the $c_{\rm G}$ term is dominant then 
$c_{\rm s}^{2} = -1/3 + {\cal O}(\Omega_{\pi}(z))$, indicating a classical instability \cite{arXiv:1008.0644}. Although we do not consider cosmological perturbations at such high redshifts in this work, the Laplace condition  is applicable in the early universe since to leading order the metric perturbations decouple then, and we can simply consider perturbations of the Galileon field evolving on the background. In this case, $c_{\rm s}^{2} = -1/3$, but even during matter domination, whenever $\rho_{\rm rad} > \rho_{\pi}$ we expect, and observe, this instability to be present, leading to the absence of viable models in 
our scans that differ substantially from the uncoupled case.  An important 
consequence is that since the 
$c_G$ term is always of order $\bar{H}^2$ larger than the $c_2$ term, 
a model with only $c_2$ and $c_G$, such as that of \cite{gubitosi} (even 
generalized to arbitrary $c_2$ so as to enable a de Sitter future state), 
is ruled out. Also, the fact that only $c_{\rm G}>0$ models survive our 
constraints implies that this term cannot dominate $\rho_{\pi}$ during 
matter and radiation domination.

\begin{figure}
\centering
\mbox{\resizebox{0.48\textwidth}{!}{\includegraphics[angle=270]{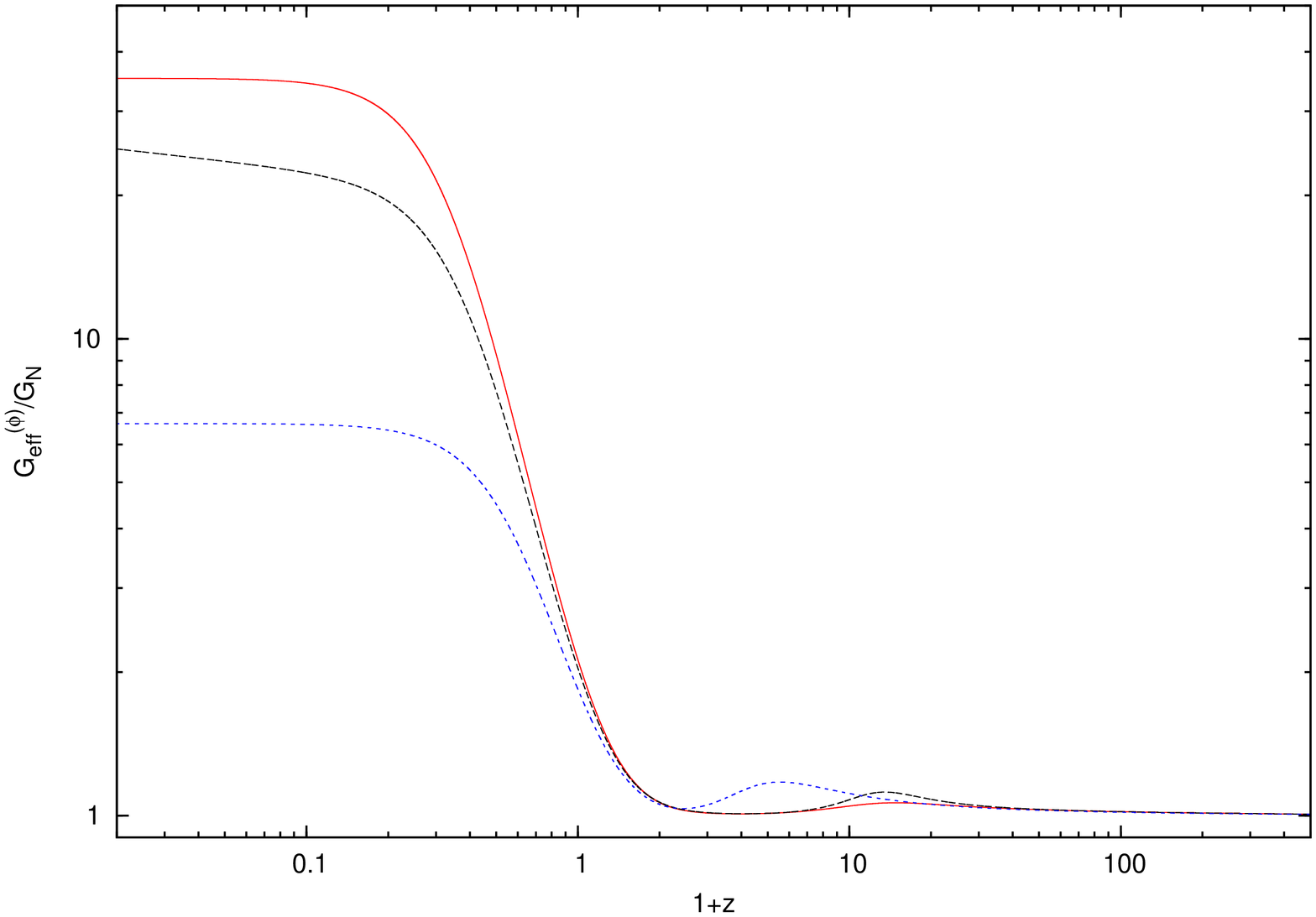}}}
\mbox{\resizebox{0.48\textwidth}{!}{\includegraphics[angle=270]{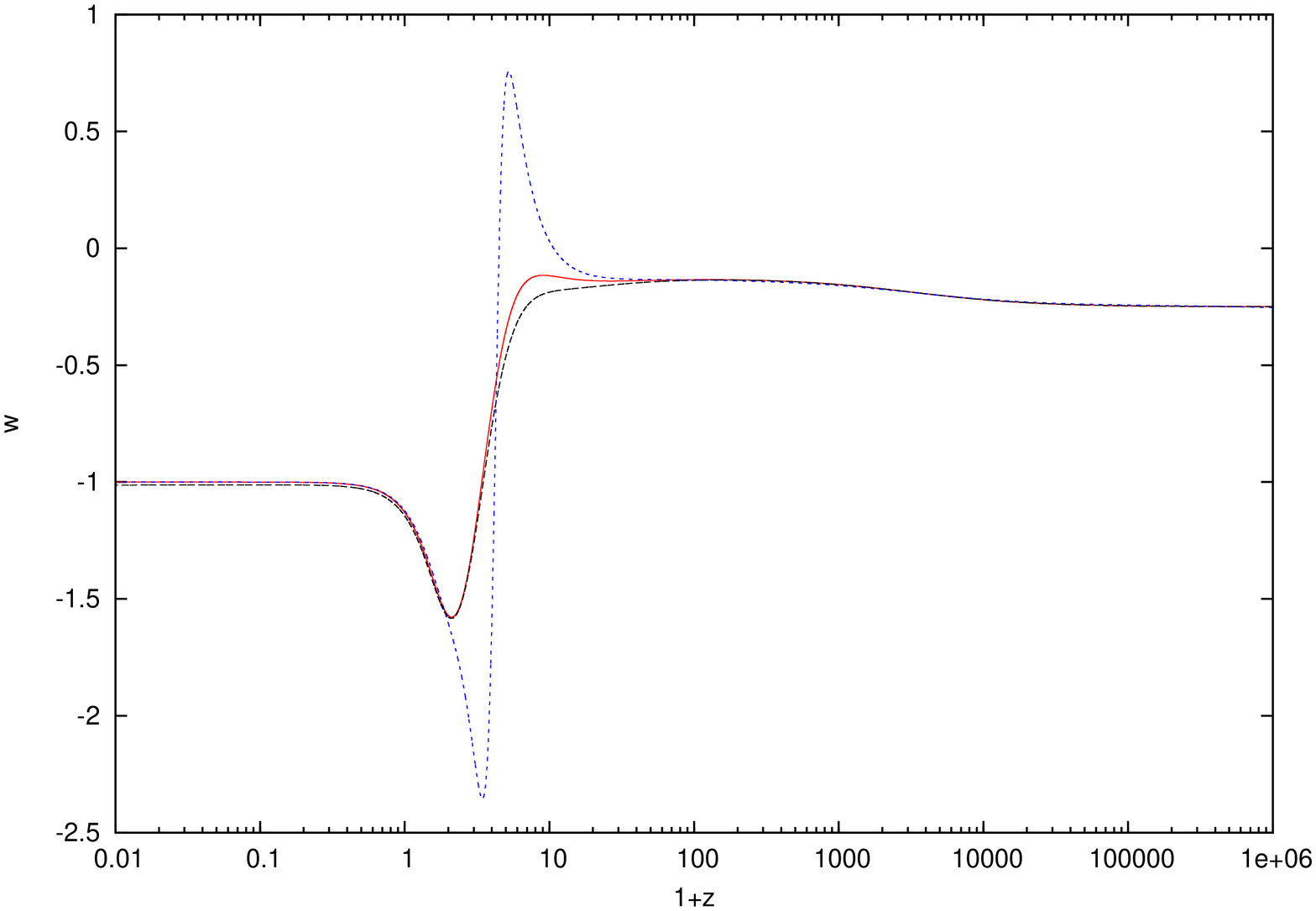}}} 
\caption{\label{fig:comp}  Typical examples of [Left panel] 
$\geff^{(\phi)}(a)$ and [Right panel] $w(a)$ for viable models in the 
uncoupled (solid red), linearly coupled  (long dash black), and derivatively 
coupled (short dash blue) classes.  For each of the cases we fix 
$c_{3} = -5.73$, $c_{4} = -1.2$ and $c_{5} = -1$, and adjust the 
parameter $c_{0}$ or $c_{\rm G}$ such that the model exists in the vicinity 
of the peak of that case's $\geff(z=1)$ histogram (specifically $c_{0} = 0.05$ 
or $c_{\rm G} = 0.4$). 
The behaviour of uncoupled and derivatively coupled models is qualitatively 
similar, with both $G_{\rm eff}^{(\phi)}$ functions exhibiting a late time 
approach to a constant positive asymptote. } 
\end{figure}

In Figure~\ref{fig:comp} we exhibit the evolution of 
$G^{(\phi)}_{\rm eff}(a)$ for a ``typical'' viable model -- that is, a model 
near the peak of the histograms for $z=1$ -- for each of the uncoupled, 
linearly coupled, and derivatively coupled cases.  We note a common 
deviation from general relativity at $z\approx10$, 
a reapproach to GR, and then the strong departure as cosmic acceleration 
occurs.  Evolving to the future beyond $z=0$, $G^{(\phi)}_{\rm eff}$ 
approaches a positive, constant value in the de Sitter asymptote for the 
derivative and uncoupled cases, while in the linear case 
$G_{\rm eff}^{(\phi)}$ will continue to grow due to the explicit dependence 
on $y$.

The nonmonotonic behaviour of $G_{\rm eff}^{(\phi)}$ at $z \approx 10$ is 
a common feature over the parameter space considered, and its origin can 
be traced to an interaction between the different $c_n$ terms.  Although 
the $c_5$ term, say, dominates at early times, the $c_{3,4}$ contributions 
grow relative to this and at some redshift they become comparable.  This 
leads to cancellations in the denominator of $\geff^{(\phi)}$ and account 
for the observed bump.  The height and redshift of this feature depend on 
the values of $c_{2-5}$ and also $\rho_{\pi}(z_i)$; for example for lower 
initial $\pi$ density, the hierarchy between the $c_{3-5}$ terms is less 
pronounced, leading to a feature at earlier times.  Generically, then, the 
paths of gravity in Galileon gravity are quite intricate and rich.

\section{Conclusions} \label{sec:concl} 

Galileon gravity is a well defined theory with a rich phenomenology. 
We explored the predictions for expansion history giving rise to 
current acceleration, despite the absence of any cosmological constant 
or potential, and found early time scaling and tracing solutions 
and late time de Sitter states.  Equally important is the behavior of 
the inhomogeneities in the Galileon field, and we discussed requirements 
for a sound theory in terms of the no-ghost condition, Laplace 
stability, and positivity of energy density. By creating a large ensemble of 
expansion histories, we were able to 
use these theoretical conditions to efficiently rule out large regions 
of the multi dimensional parameter space. In this work we focussed solely on the 
dynamics of scalar perturbations; if we include tensor modes then we expect 
even more stringent bounds on the Galileon parameters (see for example \cite{Kimura:2011qn}). Confronting the model with 
data to further constrain the parameters will be the subject of future work. 

The modifications to gravity induced by the Galileon terms can be 
characterized by gravitational strengths (effective Newton's constants) 
entering the Poisson equations for the various combinations of metric 
potentials, $\geff^{(\phi)}$, $\geff^{(\psi)}$, and $\geff^{(\phi+\psi)}$. 
We derived expressions for these and solutions in the early time and late 
time limits.  Considering special cases and general scans over the parameter 
space, we presented the ``paths of gravity'', the evolution of the effective 
gravitational strengths over cosmic history.  

Galileon models do not have simple, monotonic evolution of gravity.  One 
special feature discovered is that in the de Sitter limit the ``matter 
gravity'' $\geff^{(\psi)}$ entering into growth of structure and the 
``light gravity'' $\geff^{(\phi+\psi)}$ entering into light deflection 
are identical (unlike in many other theories), although generally different 
from general relativity.  Gravity can suddenly become stronger at some 
epoch in $1\lesssim z\lesssim 1000$, and then be restored to general 
relativity; it is intriguing to consider the possible effects on structure 
formation such as enabling early formation of massive clusters. The observed phenomenology  
of $\geff^{(\psi)}$ at intermediate redshifts will significantly increase the (linear) power of the 
matter perturbations at late times, and this will be used to severely constrain the 
uncoupled Galileon in an upcoming study.

Models which include a coupling between the Galileon field and the 
energy-momentum tensor, or equivalently the field and the Ricci scalar in 
the case of linear coupling or the field kinetic term and the Einstein 
tensor in the case of derivative coupling, are severely constrained by 
conditions for soundness of the theory.  In particular, we find that each 
coupling term must be subdominant to uncoupled Galileon terms in the 
early universe, and generally small.  Deviations from the uncoupled case 
therefore only tend to show up at late times.  For example, the linearly 
coupled case does not have a true de Sitter fixed point.  

Galileon gravity offers a rich variety in both the expansion behavior 
(e.g.\ equations of state that can track the matter component and give 
rise to early dark energy) and gravity behavior (spikes of deviations 
from, and restoration to, general relativity).  It provides a theoretically 
viable alternative to Einstein gravity, motivated by higher dimension 
geometric theories and protected by symmetries, without many of the 
naturalness issues of the cosmological constant or scalar field potentials. 
Galileon cosmology's implications for observations at high and low redshift 
make it an excellent theory to explore, and to test as an origin for cosmic 
acceleration and an extension to Einstein gravity.

\acknowledgments 

This work has been supported by World Class University grant 
R32-2009-000-10130-0 through the National Research Foundation, Ministry 
of Education, Science and Technology of Korea, and in part by the Director, 
Office of Science, Office of High Energy Physics, of the 
U.S.\ Department of Energy under Contract No.\ DE-AC02-05CH11231.  We 
thank Claudia de Rham, Giulia Gubitosi, Andrew Tolley and Michael Kopp for helpful 
discussions.

\appendix

\section{Jordan Frame}\label{sec:appT} 

In the main body of the paper we choose to work in a frame in which there 
is no direct coupling between the $\pi$ field and standard matter, instead 
the coupling is exhibited directly to the metric quantities.  To accomplish 
this we here consider the transform between the Einstein and Jordan frames 
in the weak field limit (see also \cite{arXiv:0905.1325}).    
We consider two couplings throughout this work, and we treat them 
separately here. Beginning with the linear coupling, we write  
$g_{\mu\nu} = \eta_{\mu\nu} + h_{\mu\nu}$ and the Lagrangian to quadratic 
order in $h_{\mu\nu}$ is given by 
\begin{equation} 
\label{eq:ed1}  {\cal L} = -  {M_{\rm pl}^{2} \over 2}  h^{\mu\nu} \delta G_{\mu\nu}  
+ \frac{1}{2} \sum_{i=1}^{5} c_{i} {\cal L}_{i} + {1 \over 2} h^{\mu\nu}T_{\mu\nu} + {c_{0} \over M_{\rm pl}} \pi T  \,, 
\end{equation} 
where $\delta G_{\mu\nu}$ is the linearized Einstein tensor and 
$T = \eta^{\mu\nu}T_{\mu\nu}$, where $T_{\mu\nu}$ is the energy momentum 
tensor of matter. In this linearized limit we can perform the transformation 
$h_{\mu\nu} \to \bar{h}_{\mu\nu} = h_{\mu\nu} + 2 c_{0} \pi/M_{\rm pl} \eta_{\mu\nu}$ 
to remove the $c_{0}$ coupling term. Under this transformation, neglecting 
any boundary terms, the linearized Einstein contribution in Eq.~(\ref{eq:ed1}), 
the first term on the right hand side, introduces three additional terms, 
two of which can be absorbed into the $c_{2,3}$ parameters.  In terms of 
$\bar{h}_{\mu\nu}$ and $\pi$, the Lagrangian can be written as 
\begin{equation} 
\label{eq:ed2}  {\cal L} = -  {M_{\rm pl}^{2} \over 2}  \bar{h}^{\mu\nu} 
\delta \bar{G}_{\mu\nu} +  2 c_{0} M_{\rm pl} \pi \eta^{\mu\nu} 
\delta \bar{G}_{\mu\nu}  + \frac{1}{2} \sum_{i=1}^{5} \bar{c}_{i} {\cal L}_{i} + {1 \over 2} \bar{h}^{\mu\nu}T_{\mu\nu}  \,, 
\end{equation} 
where $\bar{c}_{i}$ are the redefined constants, and 
$\delta \bar{G}_{\mu\nu}$ is the linearized Einstein tensor involving 
on $\bar{h}_{\mu\nu}$.  Having constructed the linearized Jordan frame 
limit, we promote the action to its full, nonlinear form: 
\begin{equation} 
\label{eq:ii2}   S = \int \sqrt{-g} d^{4}x \left[  \left( 1  -  2c_{0} { \pi \over M_{\rm pl}} \right)  {M_{\rm pl}^{2} R \over 2} - {c_{2} \over 2} (\partial \pi)^{2} - {c_{3} \over M^{3}}(\partial \pi)^{2} \Box \pi - {c_{4}{\cal L}_{4} \over 2} - {c_{5}{\cal L}_{5} \over 2}  - {\cal L}_{\rm m}   \right]  \ . 
\end{equation} 

Similarly, we can repeat the above procedure with the derivative 
coupling $c_{\rm G} T^{\mu\nu} \partial_{\mu}\pi\partial_{\nu}\pi$.  
Specifically, we write the linearized Lagrangian as 
\begin{equation} 
\label{eq:ed12}  {\cal L} = -  {M_{\rm pl}^{2} \over 2}  
h^{\mu\nu} \delta G_{\mu\nu}  + \frac{1}{2} \sum_{i=1}^{5} c_{i} {\cal L}_{i} + {1 \over 2} h^{\mu\nu}T_{\mu\nu} + {c_{\rm G} \over M_{\rm pl} M^{3}}  T_{\mu\nu} \partial^{\mu}\pi\partial^{\nu}\pi  \ . 
\end{equation} 
Performing the transformation $h_{\mu\nu} \to \bar{h}_{\mu\nu} = h_{\mu\nu} + 2(c_{\rm G}/M_{\rm pl}M^{3}) \partial_{\mu}\pi\partial_{\nu}\pi$, we remove the explicit coupling to matter at the expense of introducing additional mixing terms between $\bar{h}_{\mu\nu}$ and $\pi$. The transform introduces a term that can be absorbed into $c_{4}$, and a new term of the form $\partial^{\mu}\pi \partial^{\nu}\pi \delta G_{\mu\nu}$. We find 

\begin{equation} \label{eq:ed13}  {\cal L} = -  {M_{\rm pl}^{2} \over 2}  \bar{h}^{\mu\nu} \delta \bar{G}_{\mu\nu} + 2c_{\rm G} {M_{\rm pl} \over M^{3}} \partial^{\mu}\pi\partial^{\nu}\pi  \delta \bar{G}_{\mu\nu}  + \frac{1}{2} \sum_{i=1}^{5} \bar{c}_{i} {\cal L}_{i} + {1 \over 2} \bar{h}^{\mu\nu}T_{\mu\nu}  \ . 
\end{equation} 

\noindent Promoting to a full, nonlinear action, we arrive at

\begin{equation} \label{eq:append3}   
S = \int d^4 x\,\sqrt{-g} \left[ {M_{\rm pl}^{2} R \over 2} - {c_{2} \over 2} (\partial \pi)^{2} - {c_{3} \over M^{3}}(\partial \pi)^{2} \Box \pi - {c_{4}{\cal L}_{4} \over 2} - {c_{5}{\cal L}_{5} \over 2}   - {M_{\rm pl}\over M^{3}} c_{\rm G} G^{\mu\nu} \partial_{\mu}\pi\partial_{\nu}\pi  - {\cal L}_{\rm m}    \right] \,. 
\end{equation}

In regions of high curvature, the equivalence between the Jordan and Einstein frames considered here will no longer be valid, however we work with the Jordan frame quantities throughout.

\section{Field Equations} \label{sec:appfdeq} 

To obtain the Einstein and scalar field equations we vary the action of 
Eq.~($\ref{eq:ii3}$) with respect to the metric $g_{\mu\nu}$ and Galileon 
field $\pi$, obtaining 

\begin{eqnarray}  \nonumber & & {M^{3}c_{2} \over 2 } \Box \pi + c_{3}(\Box \pi)^{2} - c_{3}(\nabla_{\mu}\nabla_{\nu}\pi)^{2} - c_{3}R^{\mu\nu}\nabla_{\mu}\pi\nabla_{\nu}\pi  + M_{\rm pl} c_{\rm G} G^{\mu\nu} \nabla_{\mu}\nabla_{\nu}\pi \\  
\nonumber & & = {M_{\rm pl}M^{3}c_{0} \over 2}R  + {c_{4} \over 4M^{3}} 
\left[ -4(\Box\pi)^{3} - 8\nabla_{\alpha}\nabla_{\nu} \pi \nabla^{\nu}\nabla^{\lambda}\pi \nabla_{\lambda}\nabla^{\alpha}\pi + 12 (\Box \pi )\nabla_{\alpha}\nabla_{\beta}\pi \nabla^{\alpha}\nabla^{\beta} \pi  + 2 \Box \pi  \nabla_{\alpha}\pi \nabla^{\alpha}\pi  R \right. \\ 
\nonumber & & \left. +4 \nabla_{\alpha}\pi \nabla_{\beta}\pi \nabla^{\alpha}\nabla^{\beta} \pi R + 8 (\Box \pi) R^{\mu\nu} \nabla_{\mu}\pi \nabla_{\nu}\pi - 4 \nabla_{\alpha}\pi \nabla^{\alpha} \pi \nabla^{\lambda}\nabla^{\sigma}\pi  R_{\lambda\sigma} - 16 \nabla_{\alpha}\pi\nabla^{\alpha}\nabla^{\sigma} \pi R_{\sigma\rho} \nabla^{\rho}\pi \right. \\ 
\nonumber & & \left. - 8 \nabla_{\alpha}\pi \nabla_{\beta}\pi \nabla_{\rho} \nabla_{\sigma} \pi R^{\alpha\rho\beta \sigma} \right]  \\ 
\nonumber & &  \qquad \left(  1 - 2c_{0}{\pi \over M_{\rm pl}} \right) M_{\rm pl}^{2}G_{\mu\nu} = T_{\mu\nu} -2 M_{\rm pl} c_{0} (\nabla_{\mu}\nabla_{\nu} - g_{\mu\nu} \Box )\pi  + T_{\mu\nu}^{\rm (c_{2})} + T_{\mu\nu}^{\rm (c_{3})} + T_{\mu\nu}^{\rm (c_{4})} +T_{\mu\nu}^{\rm (c_{5})} + T_{\mu\nu}^{\rm (c_{\rm G})} 
\end{eqnarray}
where the $c_{2-5}$ contributions to the Galileon energy momentum tensor, 
$T^{(c_{2-5})}_{\mu\nu}$, were first calculated in \cite{arXiv:0901.1314}.  We repeat 
them here for completeness, and add the $c_{\rm G}$ contribution. 

\begin{eqnarray}  & & T_{\mu\nu}^{\rm (c_{2})} =  c_{2}\left[ \nabla_{\mu}\pi\nabla_{\nu}\pi - {1 \over 2} g_{\mu\nu} (\partial \pi)^{2} \right]  \\ & &  T_{\mu\nu}^{\rm (c_{3})} =   {c_{3} \over M^{3}}\left[ 2\nabla_{\mu}\pi\nabla_{\nu}\pi \Box \pi + g_{\mu\nu} \nabla_{\alpha}\pi \nabla^{\alpha}(\partial \pi)^{2} - 2 \nabla_{(\mu}\pi \nabla_{\nu)}(\partial\pi)^{2}\right]  \\ \nonumber & & T_{\mu\nu}^{\rm (c_{4})} =  -{c_{4} \over M^{6}} \left[ 4\Box \pi \nabla_{\alpha}\pi \nabla_{\mu}\pi \nabla^{\alpha}\nabla_{\nu} \pi + 4\Box \pi \nabla_{\alpha} \pi\nabla_{\nu}\pi \nabla^{\alpha}\nabla_{\mu}\pi - 2\Box\pi \Box \pi \nabla_{\mu}\pi \nabla_{\nu}\pi + 2 \Box \pi \nabla_{\alpha}\pi \nabla^{\alpha}\pi \nabla_{\mu}\nabla_{\nu} \pi  \right. \\ \nonumber & & \qquad \qquad + 4\nabla^{\alpha}\pi \nabla^{\lambda} \pi \nabla_{\alpha}\nabla_{\lambda}\pi \nabla_{\mu}\nabla_{\nu} \pi - 4 \nabla^{\alpha}\pi \nabla^{\lambda}\pi \nabla_{\alpha}\nabla_{\mu}\pi \nabla_{\lambda}\nabla_{\nu}\pi + 2 \nabla_{\mu}\pi \nabla_{\nu}\pi \nabla^{\alpha}\nabla^{\beta} \pi \nabla_{\alpha} \nabla_{\beta} \pi \\ \nonumber & & \qquad \qquad  - 2 \nabla_{\alpha}\pi \nabla^{\alpha} \pi \nabla_{\mu}\nabla_{\beta} \pi \nabla_{\nu}\nabla^{\beta} \pi - 4 \nabla_{\alpha}\pi \nabla_{\nu}\pi \nabla^{\alpha} \nabla_{\sigma}\pi \nabla^{\sigma}\nabla_{\mu}\pi - 4 \nabla_{\alpha}\pi \nabla_{\mu}\pi \nabla^{\alpha}\nabla_{\sigma}\pi \nabla^{\sigma}\nabla_{\nu} \pi  - g_{\mu\nu} (\Box\pi)^{2} \nabla_{\alpha}\pi \nabla^{\alpha} \pi \\ \nonumber & &  \qquad \qquad -4 g_{\mu\nu} \Box \pi \nabla_{\alpha}\pi \nabla_{\lambda}\pi \nabla^{\alpha}\nabla^{\lambda}\pi + 4g_{\mu\nu} \nabla_{\alpha} \pi \nabla_{\omega}\pi \nabla^{\alpha} \nabla_{\lambda}\pi \nabla^{\lambda}\nabla^{\omega}\pi + g_{\mu\nu} \nabla_{\alpha}\pi \nabla^{\alpha}\pi \nabla_{\lambda}\nabla_{\sigma}\pi \nabla^{\lambda} \nabla^{\sigma}\pi  \\ \nonumber & &  \qquad \qquad + \nabla_{\mu}\pi \nabla_{\nu}\pi \nabla_{\alpha}\pi \nabla^{\alpha}\pi R  -{g_{\mu\nu} \over 4} \nabla_{\alpha}\pi \nabla^{\alpha} \pi \nabla_{\lambda}\pi \nabla^{\lambda}\pi R - 2 \nabla_{\alpha} \pi \nabla^{\alpha} \pi  R_{\lambda \mu} \nabla_{\nu}\pi \nabla^{\lambda} \pi  - 2 \nabla_{\alpha}\pi \nabla^{\alpha}\pi R_{\lambda \nu}\nabla_{\mu}\pi \nabla^{\lambda} \pi  \\ & &  \qquad \qquad \left. + {1 \over 2} R_{\mu\nu} \nabla_{\alpha}\pi \nabla^{\alpha}\pi \nabla_{\lambda}\pi \nabla^{\lambda}\pi + 2 g_{\mu\nu} R_{\sigma \omega} \nabla^{\sigma}\pi \nabla^{\omega}\pi \nabla_{\alpha} \pi \nabla^{\alpha} \pi  - 2 \nabla_{\alpha}\pi \nabla^{\alpha}\pi g_{\mu\zeta} R^{\zeta}{}_{\sigma\nu\omega} \nabla^{\sigma}\pi \nabla^{\omega}\pi  \right]  
 \\ \nonumber & & T_{\mu\nu}^{\rm (c_{5})} =  {c_{5} \over M^{9}} \left[  \left( \Box \pi\right)^{3} \nabla_{\mu}\pi\nabla_{\nu}\pi + \left( \Box \pi \right)^{3}  \n_{\alpha}\pi \n^{\alpha}\pi g_{\mu\nu} - 3 \left(\Box \pi \right)^{2} \n_{\alpha}\pi\n^{\alpha} \pi \n_{\mu}\n_{\nu}\pi  \right.
 \\ \nonumber  & & \qquad \qquad  - 3 \left( \Box \pi\right)^{2} \n_{\rho}\pi \left( \n^{\rho} \n_{\mu}\pi \n_{\nu}\pi + \n^{\rho}\n_{\nu}\pi \n_{\mu}\pi \right) + 3 \left(\Box \pi \right)^{2} \n_{\alpha}\pi\n_{\beta}\pi \n^{\alpha}\n^{\beta} \pi g_{\mu\nu}  \\ \nonumber & & \qquad \qquad  + 6 \left( \Box \pi \right) \n_{\alpha}\pi\n^{\alpha}\pi \n_{\mu}\n_{\sigma}\pi \n^{\sigma}\n_{\nu}\pi - 6 (\Box\pi) \n^{\alpha}\pi \n^{\beta}\pi \n_{\alpha}\n_{\beta}\pi \n_{\mu}\n_{\nu}\pi - 3 (\Box\pi) \n^{\alpha}\n^{\beta}\pi \n_{\alpha}\n_{\beta} \pi \n_{\mu}\pi \n_{\nu}\pi  
  \\ \nonumber & & \qquad \qquad + 6 (\Box \pi) \n^{\alpha}\pi \n^{\beta}\pi \n_{\alpha}\n_{\mu}\pi \n_{\beta}\n_{\nu}\pi +6(\Box\pi) \n^{\alpha}\pi \n_{\alpha}\n_{\beta}\pi \left( \n^{\beta}\n_{\mu}\pi \n_{\nu}\pi + \n^{\beta}\n_{\nu}\pi \n_{\mu}\pi \right)  \\ \nonumber & & \qquad \qquad  - 3 (\Box \pi) \n_{\alpha}\pi\n^{\alpha}\pi \n_{\lambda}\n_{\sigma} \pi \n^{\lambda}\n^{\sigma}\pi g_{\mu\nu} - 6 (\Box \pi) \n^{\alpha}\pi \n^{\beta}\pi \n_{\alpha}\n_{\sigma}\pi \n_{\beta} \n^{\sigma}\pi g_{\mu\nu} - {3 \over 2} (\Box \pi) \n_{\alpha}\pi \n^{\alpha}\pi \n_{\mu}\pi \n_{\nu}\pi R  
   \\ \nonumber & & \qquad \qquad + 3 (\Box \pi) \n_{\alpha}\pi \n^{\alpha}\pi \n^{\beta}\pi \left( R_{\beta\mu}\n_{\nu}\pi + R_{\beta\nu}\n_{\mu}\pi \right)  - 3 (\Box \pi) \n_{\alpha}\pi\n^{\alpha}\pi \n^{\lambda}\pi \n^{\beta}\pi R_{\lambda\beta} g_{\mu\nu}  
    \\ \nonumber & & \qquad \qquad  + 3 (\Box \pi)  \n_{\alpha}\pi\n^{\alpha}\pi  \n^{\lambda}\pi \n^{\sigma}\pi R_{\mu\sigma\nu\lambda} + 3 \n_{\alpha}\pi\n^{\alpha}\pi \n^{\lambda}\n^{\sigma}\pi \n_{\lambda}\n_{\sigma} \pi \n_{\mu}\n_{\nu}\pi - 6  \n_{\alpha}\pi\n^{\alpha}\pi  \n_{\mu}\n^{\sigma}\pi \n_{\sigma}\n_{\lambda}\pi \n^{\lambda}\n_{\nu}\pi
     \\ \nonumber & & \qquad \qquad + 6 \n^{\alpha}\pi \n^{\beta}\pi \n_{\alpha}\n_{\beta} \pi \n^{\lambda}\n_{\mu}\pi \n_{\lambda}\n_{\nu}\pi + 6 \n^{\alpha}\pi \n^{\beta}\pi \n_{\alpha}\n_{\sigma}\pi \n^{\sigma}\n_{\beta}\pi \n_{\mu}\n_{\nu} \pi + 2 \n_{\rho}\n^{\sigma}\pi \n_{\sigma}\n^{\lambda}\pi \n_{\lambda}\n^{\rho}\pi \n_{\mu}\pi \n_{\nu}\pi 
      \\ \nonumber & & \qquad \qquad  + 3 \n^{\alpha}\n^{\beta}\pi \n_{\alpha}\n_{\beta}\pi \n_{\lambda}\pi \left( \n^{\lambda}\n_{\mu}\pi \n_{\nu}\pi + \n^{\lambda}\n_{\nu}\pi \n_{\mu}\pi \right) - 6  \n_{\rho}\pi \n^{\rho}\n^{\sigma}\pi \n_{\sigma}\n_{\lambda}\pi \left( \n^{\lambda}\n_{\mu}\pi \n_{\nu}\pi + \n^{\lambda}\n_{\nu}\pi \n_{\mu}\pi \right) 
       \\ \nonumber & & \qquad \qquad  - 6 \n^{\rho}\pi \n_{\rho}\n_{\lambda}\pi \n_{\sigma} \left( \n^{\lambda}\n_{\mu}\pi \n^{\sigma}\n_{\nu}\pi + \n^{\lambda}\n_{\nu}\pi \n^{\sigma} \n_{\mu}\pi \right) + 2  \n_{\alpha}\pi\n^{\alpha}\pi  \n_{\sigma}\n^{\lambda}\pi \n_{\lambda}\n^{\kappa}\pi \n_{\kappa}\n^{\sigma}\pi g_{\mu\nu} 
        \\ \nonumber & & \qquad \qquad  - 3 \n^{\alpha}\pi \n^{\beta}\pi \n_{\alpha}\n_{\beta}\pi \n^{\lambda}\n^{\sigma}\pi \n_{\lambda}\n_{\sigma} \pi g_{\mu\nu}  + 6 \n^{\rho}\pi \n^{\kappa}\pi \n_{\rho}\n_{\sigma}\pi \n^{\sigma}\n_{\lambda}\pi \n^{\lambda}\n_{\kappa}\pi g_{\mu\nu} 
         \\ \nonumber & & \qquad \qquad  + {3 \over 2}  \n_{\alpha}\pi\n^{\alpha}\pi \n^{\sigma}\pi \left( \n_{\sigma}\n_{\mu}\pi \n_{\nu} \pi + \n_{\sigma}\n_{\nu} \pi \n_{\mu}\pi \right) R - {3 \over 2}  \n_{\alpha}\pi\n^{\alpha}\pi \n^{\lambda}\pi \n^{\sigma}\pi \n_{\lambda}\n_{\sigma}\pi R g_{\mu\nu} 
          \\ \nonumber & & \qquad \qquad  + 3  \n_{\alpha}\pi\n^{\alpha}\pi  \n^{\lambda}\pi \n^{\sigma}\pi \n_{\lambda}\n_{\sigma}\pi R_{\mu\nu} + 3  \n_{\alpha}\pi\n^{\alpha}\pi \n^{\lambda}\pi \n^{\sigma}\pi R_{\lambda\sigma} \n_{\mu}\n_{\nu}\pi + 3  \n_{\alpha}\pi\n^{\alpha}\pi \n^{\lambda}\n^{\sigma}\pi R_{\lambda\sigma} \n_{\mu}\pi \n_{\nu}\pi 
           \\ \nonumber & & \qquad \qquad  - 3  \n_{\alpha}\pi\n^{\alpha}\pi  \n^{\lambda}\pi \n_{\lambda}\n^{\sigma}\pi \left( R_{\sigma\mu} \n_{\nu}\pi + R_{\sigma\nu} \n_{\mu}\pi \right) - 3  \n_{\alpha}\pi\n^{\alpha}\pi  \n^{\lambda}\pi \n^{\sigma}\pi \left( R_{\lambda\mu}\n_{\nu}\n_{\sigma}\pi + R_{\lambda\nu}\n_{\mu}\n_{\sigma}\pi \right)  
            \\ \nonumber & & \qquad \qquad  - 3  \n_{\alpha}\pi\n^{\alpha}\pi  \n^{\sigma} R_{\sigma\lambda} \left( \n^{\lambda}\n_{\mu}\pi \n_{\nu}\pi + \n^{\lambda}\n^{\nu}\pi \n_{\mu} \pi \right) + 6  \n_{\alpha}\pi\n^{\alpha}\pi  \n^{\sigma}\pi \n_{\sigma}\n^{\lambda}\pi R_{\lambda\kappa} \n^{\kappa}\pi g_{\mu\nu} 
             \\ \nonumber & & \qquad \qquad - 3  \n_{\alpha}\pi\n^{\alpha}\pi  \n^{\sigma}\pi\n^{\lambda}\n^{\kappa}\pi \left( R_{\mu\lambda\sigma\kappa}\n_{\nu}\pi + R_{\nu\lambda\sigma\kappa}\n_{\mu}\pi \right) + 3  \n_{\alpha}\pi\n^{\alpha}\pi  \n^{\sigma}\pi \n^{\lambda}\pi \left( R_{\mu\sigma\lambda\kappa} \n^{\kappa}\n_{\nu}\pi + R_{\nu\sigma\lambda\kappa}\n^{\kappa}\n_{\mu}\pi \right)
              \\  & & \qquad \qquad \left. - 3  \n_{\alpha}\pi\n^{\alpha}\pi  \n^{\sigma}\pi \n_{\sigma}\n^{\lambda} \pi\n^{\kappa}\pi \left( R_{\mu\lambda\nu\kappa} + R_{\nu\lambda\mu\kappa} \right) + 3  \n_{\alpha}\pi\n^{\alpha}\pi  \n^{\sigma}\pi \n^{\lambda}\pi \n^{\kappa}\n^{\tau}\pi R_{\sigma\kappa\lambda\tau} g_{\mu\nu} \right]
\\ & & T_{\mu\nu}^{\rm (c_{\rm G})} =   {M_{\rm pl} \over M^{3}} c_{\rm G} \left[ g_{\mu\nu} \Box\pi \Box\pi - 2 \Box\pi \nabla_{\mu}\nabla_{\nu}\pi + 2\nabla_{\mu}\nabla_{\lambda}\pi \nabla_{\nu}\nabla^{\lambda}\pi - g_{\mu\nu} \nabla_{\lambda}\nabla_{\alpha}\pi \nabla^{\lambda}\nabla^{\alpha}\pi \right] \\ \nonumber & & \qquad \qquad \qquad - {M_{\rm pl} \over M^{3}} c_{\rm G} \left[ R_{\mu\nu}\nabla_{\alpha}\pi \nabla^{\alpha}\pi + R \nabla_{\mu}\pi \nabla_{\nu}\pi - {1 \over 2} g_{\mu\nu} R \nabla_{\alpha}\pi \nabla^{\alpha}\pi \right]  \\ \nonumber & & \qquad \qquad \qquad   + 2 {M_{\rm pl} \over M^{3}}c_{\rm G} \left[ R_{\lambda \nu}\nabla^{\lambda}\pi \nabla_{\mu} \pi  + R_{\lambda\mu} \nabla^{\lambda}\pi \nabla_{\nu}\pi - g_{\mu\nu} R_{\rho\lambda}\nabla^{\rho}\pi \nabla^{\lambda}\pi + R^{\sigma}{}_{\mu\beta\nu} \nabla^{\beta}\pi \nabla_{\sigma}\pi \right]  \ . 
\end{eqnarray}

\section{Perturbed Field Equations} \label{sec:applineq} 

To obtain the perturbed field equations in the subhorizon limit, taking 
the Newtonian gauge and defining $\delta y = \delta \pi /M_{\rm pl}$, 
we linearize the perturbed $(i, j \neq i)$ and $(0,0)$ Einstein, $\pi$ dynamical, 
and fluid conservation equations in $\delta y$, the metric potentials $\phi$ 
and $\psi$, and the density perturbation $\delta_{\rm m}$.  We obtain the 
following equations 
\begin{eqnarray} 
\nonumber & & -\left[  1 - 2c_{0}y\right] \left(\bar{\nabla}^{2} \psi - \bar{\nabla}^{2} \phi \right) = -2 c_{0} \bar{\nabla}^{2}\delta y \\  \nonumber & &   \hspace{30mm} - c_{4} \left[ 6\bar{H}^{3} x^{2} \left( \bar{H}' x + \bar{H} x' + {1 \over 3}  \bar{H} x \right) \bar{\nabla}^{2} \delta y + {\bar{H}^{4} x^{4} \over 2} \bar{\nabla}^{2} \phi + {3 \over 2} \bar{H}^{4}x^{4} \bar{\nabla}^{2} \psi \right] \\  \nonumber & & \hspace{30mm} + c_{\rm G} \left[ 2\left( \bar{H}\bar{H}'x + \bar{H}^{2} x' + \bar{H}^{2}x \right) \bar{\nabla}^{2} \delta y + \bar{H}^{2} x^{2} \bar{\nabla}^{2} \phi +  \bar{H}^{2} x^{2} \bar{\nabla}^{2} \psi \right] \\ \label{eq:n1} & &  \hspace{30mm} + c_{5} \left[   3\bar{H}^{6} x^{5} \bar{\nabla}^{2} \psi - 3\bar{H}^{5}x^{4} \left( \bar{H}x' + \bar{H}' x \right) \bar{\nabla}^{2} \phi + \bar{H}^{5} x^{3} \left( 12 \bar{H} x'  + 15 \bar{H}' x + 3\bar{H} x \right) \bar{\nabla}^{2} \delta y   \right] \\ \nonumber & & \\  \nonumber & &    {2 \over a^{2}}\left[ 1  - 2c_{0} y\right] \bar{\nabla}^{2} \phi =   {\rho_{\rm m} \over H_{0}^{2} M_{\rm pl}^{2}}\delta_{\rm m} + \left( 2 c_{3} \bar{H}^{2} {x^{2} \over a^{2}} - {2 c_{0}  \over a^{2}}\right) \bar{\nabla}^{2}\delta y  - c_{4} \left( {12\bar{H}^{4}x^{3} \over a^{2}} \bar{\nabla}^{2}\delta y - {3 \bar{H}^{4} x^{4}\over a^{2}} \bar{\nabla}^{2} \phi  \right) \\ \label{eq:n2} & & \hspace{10mm}  + c_{\rm G} \left[ 4\bar{H}^{2}{x \over a^{2}} \bar{\nabla}^{2} \delta y - 2\bar{H}^{2} {x^{2} \over a^{2}} \bar{\nabla}^{2} \phi \right]  + c_{5} \left[ {15 \bar{H}^{6} x^{4} \over a^{2}} \bar{\nabla}^{2} \delta y - {6 \bar{H}^{6}x^{5} \over a^{2}} \bar{\nabla}^{2} \phi \right] \\ \nonumber & & \\ \nonumber  & & -2 \left[  c_{3} \bar{H}\bar{H}' x  + c_{3}\bar{H}^{2} x'  + 2c_{3}\bar{H}^{2} x - {c_{2} \over 4}\right] \bar{\nabla}^{2} \delta y - c_{3}\bar{H}^{2} x^{2}  \bar{\nabla}^{2}\psi = -c_{0}  \left[ \bar{\nabla}^{2} \psi - 2 \bar{\nabla}^{2} \phi \right]  + c_{4} \left[   - 6 \bar{H}^{4}x^{3} \bar{\nabla}^{2}\psi  \right.  \\ \nonumber  & &   \left. + \left( 6\bar{H}^{3}x^{2}\left( \bar{H}x' + x\bar{H}' \right) + 2 \bar{H}^{4} x^{3} \right) \bar{\nabla}^{2} \phi  - \left( 12 \bar{H}^{3}x\left( \bar{H}x' + x\bar{H}' \right) + 13 \bar{H}^{4} x^{2} + 6\bar{H}^{3}x^{2} \bar{H}'  \right) \bar{\nabla}^{2}\delta y \right] \\ \nonumber & & + c_{\rm G} \left( 2\bar{H}\bar{H}' + 3\bar{H}^{2} \right) \bar{\nabla}^{2} \delta y  + 2 c_{\rm G} \bar{H}^{2}x \bar{\nabla}^{2} \psi  - 2 c_{\rm G} \left( \bar{H} \left( \bar{H} x' + \bar{H}' x \right) + \bar{H}^{2} x \right) \bar{\nabla}^{2} \phi  \\ \label{eq:n3} & & + c_{5} \left[  \left( 18\bar{H}^{6}x^{2}x' + 30\bar{H}^{5}x^{3}\bar{H}' + 12\bar{H}^{6}x^{3} \right)\bar{\nabla}^{2}\delta y + {15 \over 2} \bar{H}^{6} x^{4}\bar{\nabla}^{2}\psi - \left( 3\bar{H}^{6}x^{4} + 15\bar{H}^{5}x^{4} \bar{H}' + 12 \bar{H}^{6}x^{3}x' \right) \bar{\nabla}^{2}\phi  \right]  \\ \nonumber & &     \\ \label{eq:n4} & & \bar{H}^{2} \delta''_{\rm m} + \bar{H} \bar{H}' \delta'_{\rm m} +2 \bar{H}^{2} \delta'_{\rm m} =  {1 \over a^{2}} \bar{\nabla}^{2} \psi 
\end{eqnarray} 
where $\bar{\nabla} = \nabla/H_{0}$, $\rho_m$ is the matter density, 
and $\delta_m=\delta\rho_m/\rho_m$.  We have assumed that the 
quasistatic approximation holds at all times: 
$\delta \ddot{y}, \ddot{\phi}, \ddot{\psi} \ll \bar{\nabla}^{2} \delta y,\bar{\nabla}^{2}  \phi,\bar{\nabla}^{2} \psi$.  Note that in Eq.~(\ref{eq:n1}) the ``slip'' between 
$\phi$ and $\psi$ is not sourced by $c_2$ or $c_3$. The above equations are in agreement with those derived in \cite{arXiv:1011.6132} in the uncoupled case $c_{0} = c_{\rm G} = 0$.

Equations ($\ref{eq:n1}$) and ($\ref{eq:n3}$) can be used to eliminate 
$\bar{\nabla}^{2}\delta y$ and $\bar{\nabla}^{2}\psi$  from the 
Poisson equation~($\ref{eq:n2}$). Hence we can describe the effect of 
the Galileon on subhorizon density perturbations through two functions 
of the background quantities $\bar{H},x,y$.  We define these functions 
$G_{\rm eff}^{(\phi)}$ and $G_{\rm eff}^{(\psi)}$ in 
Eqs.~(\ref{eq:geffc4}) and (\ref{eq:geffc5}), and 
$G_{\rm eff}^{(\phi + \psi)} = [\geff^{(\phi)}+\geff^{(\psi)}]/2$.

\section{No-Ghost Condition in Detail} \label{sec:appghost} 

For a given coupling, the action ($\ref{eq:ii3}$) describes a model 
with five free parameters: the magnitude of the coupling strength 
and the kinetic terms $(c_{2},c_{3},c_{4},c_{5})$. However, one must be careful to restrict our analysis to the parameter space in which the model is theoretically viable. Of particular importance is the absence of propagating ghost degrees of freedom, that is degrees of freedom whose kinetic contribution to the Hamiltonian is negative, making it unbounded from below. 

In \cite{arXiv:1005.0868} a ``no-ghost'' condition was derived for models 
of the form 
\begin{equation} 
\label{eq:ip2} S = \int d^{4}x \sqrt{-g} \left[ {1 \over 2} F(\pi) R + {1 \over 2} f_{2}(\pi,X) + \zeta(\pi)(\partial\pi)^{2}\Box\pi \right]  \,, 
\end{equation} 
where $X = -g^{\mu\nu} \partial_{\mu}\pi\partial_{\nu}\pi/2$, and $F(\pi)$, $f_{2}(\pi,X)$ and $\zeta(\pi)$ are arbitrary functions. The no-ghost condition corresponds to  \cite{arXiv:1005.0868}
\begin{equation} \label{eq:ghostf} \left( 24\zeta H \dot{\pi} - 8\zeta_{,\pi} \dot{\pi}^{2} + f_{2,X} + f_{2,XX}\dot{\pi}^{2} \right) F \dot{\pi}^{2} + 3\left( \dot{F} - 2\zeta \dot{\pi}^{3} \right)^{2} > 0 \end{equation}

The action ($\ref{eq:ip2}$)  is equivalent to ours with 
$c_{4} = c_{5} = c_{0} = c_{\rm G} = 0$; in this case we can use the above 
condition as a constraint on our parameter space. When we introduce the 
couplings $c_{0,G}$ and more general kinetic terms $c_{4,5}$ the analysis 
must be redone as follows. 

To construct the no-ghost condition, we utilise four equations: the 
$\pi$ equation of motion and the $(0,0)$, trace, and $(i,j\ne i)$ Einstein 
equations. To deduce whether any propagating degrees of freedom exhibit 
ghost behaviour, it suffices to look for the sign of the coefficient of 
the $\ddot{\phi}$ and $\delta \ddot{y}$ terms. Our approach is to note that 
the $(i,j\ne i)$ equation gives an algebraic relation between $\delta y$, 
$\psi$ and $\phi$, and the $(0,0)$ Einstein equation contains only 
$\nabla^{2}\psi$, $\nabla^{2}\phi$ and $\nabla^{2}\delta y$ terms (and first 
order time derivatives, which are not important in what follows).  
Therefore combining these equations yields a relationship between 
$\nabla^{2} \phi$ and $\nabla^{2} \delta y$. The $\pi$ equation of motion 
and trace Einstein equation contain the terms $\delta \ddot{y}$, 
$\ddot{\phi}$, $\nabla^{2}\psi$, $\nabla^{2} \phi$ and 
$\nabla^{2} \delta y$, and therefore by eliminating $\nabla^{2} \psi$ 
and using our relationship between $\nabla^{2}\delta y$ and 
$\nabla^{2} \phi$ we can construct equations of the form 

\begin{equation}  
{\bf A}  \ddot{{\bf x}} + {\bf B} \nabla^{2} {\bf x}  = {\bf C} \ . 
\end{equation} 
Here ${\bf A}, {\bf B}$ are diagonal $2 \times 2$ matrices 
that are functions of background quantities such as $\ddot{\pi}$,  
$H$ etc.\ 
and ${\bf x}$ is the two dimensional vector containing $\delta y$ 
and $\phi$. The matrix ${\bf C}$ contains lower order time and 
spatial derivatives of ${\bf x}$ and is unimportant in calculating 
the no-ghost condition. For the simple case of a minimally coupled, 
canonical scalar field containing the $c_{2}$ term only, the no-ghost 
condition corresponds to $c_{2} > 0$, and therefore we simply need to 
ensure that ${\bf A}$ preserves the same sign conventions. 

We begin with a simple case taking $c_{2,3}$ only. The $(i,j\ne i)$ 
equation gives $\psi = \phi$, and the $(0,0)$ component of the 
Einstein equations yields the relationship 

\begin{equation}  \nabla^{2} \phi  = {c_{3} \over M_{\rm pl}^{2} M^{3}} (\dot{\pi})^{2} \nabla^{2} \delta y+ \dots 
\end{equation} 
which will not be necessary in what follows.  The $\dots$ 
denote lower derivative contributions to the equations 
that will not be relevant to the stability condition derived here. 

The trace and $\pi$ equations are

\begin{eqnarray} & &  \left( 6\ddot{\phi} - 2\nabla^{2}\phi \right) = {c_{3} \over  M_{\rm pl} M^{3}} \left( 6\dot{\pi}^{2} \delta \ddot{y} - 2\dot{\pi}^{2} \nabla^{2} \delta y \right) + \dots \\ 
& & {M^{3} c_{2} \over 2} \left( -\delta \ddot{y} + \nabla^{2} \delta y \right) + 6c_{3} H \dot{\pi} \delta \ddot{y} - 2c_{3} \left( \ddot{\pi} + 2H\dot{\pi} \right) \nabla^{2} \delta y - c_{3} \dot{\pi}^{2} \left( 3\ddot{\phi} - \nabla^{2} \phi \right)+ \dots = 0 
\end{eqnarray} 
Eliminating $\phi$, we find that the coefficient of the $\delta \ddot{y}$ term is given by

\begin{equation}\label{eq:gh1} -{M^{3} c_{2} \over 2} + 6c_{3} H \dot{\pi} - 3c_{3}^{2}  {\dot{\pi}^{4} \over M_{\rm pl}^{2} M^{3}} < 0 \end{equation} 

\noindent which must be less than zero to ensure that perturbations of the $\pi$ field have the correct sign. This condition is in agreement with Eq.~($\ref{eq:ghostf}$).

We now expand our approach to include the $c_{4,5,G}$ terms in the 
action. Performing exactly the same steps as above, we find the 
following inequality for the no-ghost condition 
\begin{equation}  
\label{eq:ghost} -{c_{2} \over 2} + 6c_{3} \bar{H}^{2}x + 3c_{\rm G}\bar{H}^{2} - 27c_{4} \bar{H}^{4} x^{2} + 30 c_{5} \bar{H}^{6}x^{3} +    2{\left( 3c_{3}\bar{H}^{2}x^{2} + 6c_{\rm G}\bar{H}^{2}x - 18c_{4} \bar{H}^{4} x^{3}  + {45 \over 2} c_{5} \bar{H}^{6} x^{4}  -3c_{0} \right)^{2} \over -6( 1 -2c_{0}y) - 6c_{\rm G} \bar{H}^{2}x^{2} + 9c_{4} \bar{H}^{4} x^{4} - 18c_{5} \bar{H}^{6} x^{5}} < 0 
\end{equation} 
This must be satisfied at all times during the cosmological evolution 
to ensure that at the level of linear perturbations around the 
cosmological background $\delta \pi$ has a kinetic term that 
contributes positively to the Hamiltonian. Note that the no-ghost 
condition ($\ref{eq:ghost}$) can be written in terms of the 
$\kappa_{i}$ functions as 

\begin{equation} 
\kappa_{2} + {3 \over 2} {\kappa_{5}^{2} \over \kappa_{4}} < 0 \ .
\end{equation}

Returning to our solutions for the linear coupling case, 
Eqs.~(\ref{eq:l10})-(\ref{eq:l2}) valid during matter domination, we 
can show that for parameters that give a positive energy density 
$\rho_{\pi} > 0$, the condition ($\ref{eq:ghost}$) is not satisfied. 
To see this, we note that at early times the no-ghost condition is 
well approximated by the first term $\kappa_2$, 
\begin{equation}  
\label{eq:ghost2} -{c_{2} \over 2} + 6c_{3} \bar{H}^{2}x + 3c_{\rm G}\bar{H}^{2} - 27c_{4} \bar{H}^{4} x^{2} + 30 c_{5} \bar{H}^{6}x^{3}   < 0 \ . 
\end{equation}

Consider the various dominating pairs, starting with $c_5, c_0$. 
To ensure that $\rho_{\pi} > 0$, we are forced to take $c_{0} > 0$ 
as $A_{0}$ must be real.  This in turn forces $c_5>0$, which 
violates Eq.~($\ref{eq:ghost2}$).  We find the same dilemma for the 
 $c_4$ solution and so on until 
we reach $c_2$.  Note that we have not shown that a ghost is generically present for the linearly coupled model whenever $\rho_{\pi} > 0$, but rather just for the particular solutions found in \ref{sec:lin}.  In fact during radiation domination the no-ghost condition is generically satisfied, as in the uncoupled case.

The no-ghost condition derived here is applicable to linear 
perturbations on an FRW background. A more complete analysis 
should be undertaken in regimes where the non-linear nature of 
the derivative self couplings becomes significant.

\section{Laplace Instability} \label{sec:applapl} 

In the previous section we considered the sign of the $\delta \ddot{y}$ term 
in the linearized perturbation equations, and how the no-ghost condition can 
be used to place constraints on the Galileon parameters. In this section we 
consider the existence of a second instability that must also be 
avoided -- the Laplace instability \cite{arXiv:1008.4236} (see also \cite{arXiv:1110.3878}
for a very complete discussion of instabilities). This is a condition on 
the sound speed of the $\pi$ field perturbation, obtained by linearizing 
the Einstein and $\pi$ equations and eliminating the metric potentials to 
obtain a wave-like equation for $\delta \pi$. Specifically, we combine the $(0,0)$, 
trace, and $(i, j\neq i)$ Einstein equations with the perturbed $\pi$ equation 
of motion, keeping all terms containing second order time and spatial 
derivatives of the fields. We find the following expression for $\delta \pi$: 

\begin{equation} \left(\kappa_{2} + {3 \over 2} {\kappa_{5}^{2} \over \kappa_{4}} \right) \delta \ddot{\pi} + \left( {   2\kappa_{3}\kappa_{5}^{2}   +2 \kappa_{4}^{2}\kappa_{6} - 4\kappa_{1} \kappa_{4}\kappa_{5}  \over 2 \kappa_{4}^{2}}\right) \nabla^{2}\delta \pi = \dots 
\end{equation} 
Again, $\dots$ represents terms unimportant to the stability argument. 
The negative definiteness of the $\delta\ddot\pi$ coefficient is the 
no-ghost condition, and the negative definiteness of the ratio of the 
$\nabla^2\delta\pi$ and $\delta\ddot\pi$ coefficients is the Laplacian 
stability condition.  This ratio is the negative of the sound speed squared, 
\begin{equation} \label{eq:p1}
c_{s}^{2} \equiv {  4\kappa_{1} \kappa_{4}\kappa_{5} -  2\kappa_{3}\kappa_{5}^{2}  - 2 \kappa_{4}^{2}\kappa_{6}   \over  \kappa_{4} \left( 2\kappa_{4}\kappa_{2} + 3 \kappa_{5}^{2}\right)  } \ . 
\end{equation} 
The Laplace stability condition is $c_s^2\ge0$.   If this condition is violated, imaginary frequency behaviour would follow, leading to exponential growth of the $\pi$ field perturbation. 
(It is conceivable that such behaviour is benign if the timescale 
associated with the growth is suitably large, however we opt for a 
conservative approach and enforce $c_s^2\ge0$ at all times during 
the cosmological evolution.) 

At early times when $\Omega_\pi\ll 1$, the sound speed simplifies 
to $c_s^2=-\kappa_6/\kappa_2$.  At late times, for the uncoupled case 
we can derive a relatively straightforward 
asymptotic de Sitter form for the sound speed, corresponding to 
\begin{equation} 
 { \left( 14 + 3C - 4D \right) \left( 416 + 108C + 9C^{2} - 32D - 16D^{2}\right) \over 54 \left(8 + 3C - 4D \right) \left( 20 + 3C - 4D \right) } 
\geq 0 \label{eq:unclaplace} 
\end{equation} 
where $C$ and $D$ are defined in Eqs.~(\ref{eq:C1}), (\ref{eq:D1}). 

In the main body of the text, we impose the positivity of $\rho_{\pi}$ 
and the no-ghost and Laplace conditions for all $z < 500$. There is an 
additional condition that one might impose, that the $\delta \pi$ field 
must avoid ``superluminal'' behaviour $c_s^{2} > 1$.  However, if such a 
condition were violated it is not necessarily the case that the model 
is ruled out; see for example \cite{arXiv:0708.0561} for a detailed discussion 
of superluminal propagation in a different class of scalar fields. 
In addition, a more complete analysis should take into account the full perturbation 
equations when constructing the scalar field dispersion relation, including effective mass terms 
of the form $\bar{H}^{2} \delta y$. For these reasons we note the potential existence 
of this constraint but do not impose it when scanning over the parameter 
space for viable models in Sec.~\ref{sec:paths}.


\end{document}